\documentstyle[12pt]{article}   
\begin{document}
\newcommand{\beqn}{\begin{eqnarray}}   
\newcommand{\eeqn}{\end{eqnarray}}   
\newcommand{\be}{\begin{equation}}   
\newcommand{\ee}{\end{equation}}   
\newcommand{\ba}{\begin{array}}   
\newcommand{\ea}{\end{array}}   
\newcommand{\R}{{\rm\bf R}}  
\newcommand{\Z}{{\rm\bf Z}}   
\newcommand{\C}{{\rm\bf C}}
  \newcommand{\N}{{\rm\bf N}}  
\newcommand{\T}{{\rm\bf T}}   
\newcommand{\pa}{\partial}   
\newcommand{\re}{\ref}   
\newcommand{\ci}{\cite}   
\newcommand{\la}{\label}   
\newcommand{\bfr}{\begin{flushright}}   
\newcommand{\efr}{\end{flushright}}   
\newcommand{\bfl}{\begin{flushleft}}   
\newcommand{\efl}{\end{flushleft}}   
\newcommand{\fr}{\frac}   
\newcommand{\ov}{\overline}   
\newcommand{\ve}{\varepsilon}   
\newcommand{\de}{\delta}   
\newcommand{\al}{\alpha}   
\newcommand{\ga}{\gamma}\newcommand{\Ga}{\Gamma}   
\newcommand{\si}{\sigma}   
\newcommand{\ds}{\displaystyle}   
\newcommand{\pr}{\prime}   
\newcommand{\La}{\Lambda}   
\newcommand{\Lr}{\Longrightarrow}   
\newcommand{\De}{\Delta}   
\newcommand{\Si}{\Sigma}   
\newcommand{\ti}{\tilde}   
\newcommand{\Om}{\Omega}   
\newcommand{\om}{\omega}   
\newcommand{\na}{\nabla}   
\newcommand{\lam}{\lambda}   
\newcommand{\Lam}{\Lambda}   
\newcommand{\br}{|\kern-.25em|\kern-.25em|}   
\renewcommand{\theequation}{\thesection.\arabic{equation}}   
\def\n#1{\vert #1 \vert}                                    
\def\nn#1{\Vert #1 \Vert}                                   
\def\Re {{\rm Re\, }}                                       
\def\Im {{\rm Im\,}}                                        
\newcommand{\const}{\mathop{\rm const}\nolimits}   
\newcommand{\supp}{\mathop{\rm supp}\nolimits}   
\newcommand{\diam}{\mathop{\rm diam}\nolimits}
\newtheorem{theorem}{Theorem}[section]   
\renewcommand{\thetheorem}{\arabic{section}.\arabic{theorem}}   
\newtheorem{definition}[theorem]{Definition}   
\newtheorem{deflem}[theorem]{Definition and Lemma}   
\newtheorem{lemma}[theorem]{Lemma}   
\newtheorem{example}[theorem]{Example}   
\newtheorem{remark}[theorem]{Remark}   
\newtheorem{remarks}[theorem]{Remarks}   
\newtheorem{cor}[theorem]{Corollary}   
\newtheorem{pro}[theorem]{Proposition}

\begin{titlepage}
\hspace{5cm} Submitted to {\em Journal of Mathematical Physics}, 2002 
\vspace{5mm} 

 \begin{center}   
 {\Large\bf  On the Convergence to Statistical 
Equilibrium    
\bigskip\\   
for Harmonic Crystals}\\   
 \vspace{2cm}   
{\large T.V.Dudnikova   
\footnote{Supported partly by the 
 START project 
"Nonlinear  Schr\"odinger and Quantum Boltzmann Equations'' (FWF Y 137- TEC) 
of N.J.Mauser 
 and  
research grants of DFG (436 RUS 113/615/0-1) and RFBR (01-01-04002).}   
}\\   
{\it   
M.V.Keldysh Institute\\   
of Applied Mathematics RAS\\    
 Moscow 125047, Russia}\\   
e-mail:~dudnik@elsite.ru, dudnik@mat.univie.ac.at   
\bigskip\\   
{\large A.I.Komech$^{1,}\!\!$   
\footnote{ On leave Department of Mechanics and Mathematics,  
Moscow State University, Moscow 119899, Russia. 
Supported partly by   
Max-Planck Institute for the Mathematics in Sciences 
(Leipzig)  
.}   
}\\   
{\it Institute of Mathematics \\ Vienna University\\   
Vienna A-1090, Austria}\\   
e-mail:~komech@mat.univie.ac.at   
\bigskip\\   
 {\large H.~Spohn}\\   
{\it Zentrum Mathematik\\   
Technische Universit\"at\\   
M\"unchen D-80290, Germany}\\   
 e-mail:~spohn@mathematik.tu-muenchen.de   
\end{center}   
 \vspace{1cm}   
   
 \begin{abstract}   
We consider the dynamics of   
a harmonic crystal in $d$ dimensions with $n$ components, 
$d,n$ arbitrary,  $d,n \ge 1$,   
and study the   
distribution $\mu_t$   
 of the solution at  time $t\in\R$.   
The initial measure $\mu_0$    has   
a  translation-invariant  correlation  matrix, 
zero mean,   
 and finite mean energy density. It  also   
satisfies a   
 Rosenblatt- resp. 
Ibragimov-Linnik type mixing condition.   
 The main result is the convergence of $\mu_t$ to   
 a Gaussian measure as $t\to\infty$.   
The proof is based on   
the long time asymptotics   
 of the Green's function and   
on Bernstein's ``room-corridors''  method.   
 \end{abstract}   
\end{titlepage}   
   
 \section{Introduction}   
Despite considerable efforts, the convergence to equilibrium 
for a mechanical system has remained as an extremely 
difficult problem. 
It has been recognized early on that for an infinitely 
extended system, possibly 
on top of local hyperbolicity, the flow of statistical 
information  
to infinity serves as a mechanism for relaxation. 
The two prime examples are the ideal gas and the 
harmonic crystal. 
We consider here the latter case. 
In the harmonic approximation the crystal is 
characterized by the  
displacement field $u(x)$, where $x\in\Ga$, $\Ga$ is 
a regular lattice in  
$\R^d$, and $u(x)\in\R^n$ with $n$ depending on 
the number of atoms 
in the unit cell. The field  $u(x)$ is governed 
by a discrete wave equation. 
We will consider arbitrary $d,n$ and for notational simplicity 
set $\Ga=\Z^d$. 
 
Our motivation to return to a well studied model is to   
a much wider class of initial measures than before. This project 
requires novel mathematical techniques. They have been 
 developed for the 
wave and Klein-Gordon equation on $\R^d$ in \cite{DKKS} - \ci{DKS}, 
but the discrete structure poses extra difficulties. 
 
Let us briefly comment on previous work. In  
\cite{LL}   
a general criterion is given which ensures mixing   
and Bernoulliness   
of the corresponding mechanical flow. 
Thereby the convergence to equilibrium is  
established   
for initial measures which are absolutely   
continuous with respect to the canonical Gaussian measure.  
In  
\cite{LL}   
moments of the displacement field are studied. 
This allows to reduce the spectral analysis of the 
Liouvillean flow  
to the spectral properties of the dynamical group 
defined on solutions of finite energy. Since the crystal 
is assumed  
to be homogeneous, these spectral properties are determined 
by the dispersion 
relations $\om_k(\theta)$, $k=1,...,n$. The Liouvillean 
flow is mixing and even Bernoulli,   
if, 
except for crossing points,
each   
$\om_k(\theta)$ is a real-analytic function which 
is not   
identically constant. In particular, 
the Lebesgue measure of the set   
$\{\theta\in \T^d:\,\nabla\om_k(\theta)=0\}$ is equal to 
zero. 
In \cite{SL}, for the case   $d=n=1$, initial measures  
are considered which have distinct temperatures to the 
left 
and to the right. 
In \cite{BPT}, again $d=n=1$, the convergence to 
equilibrium is proved  
for a more general class of initial measures  
characterized by
 a mixing condition of   
 Rosenblatt- resp. Ibragimov-Linnik type and which are  
asymptotically translation-invariant  to the left and to the 
right.

The detailed stationary phase analysis of \cite{BPT} does not  
directly generalize to $d\ge 2$. Rather we have to develop 
a novel `cutoff strategy' which more carefully exploits  
the mixing condition in Fourier space. 
This approach allows us to  all $d$ 
with in essence the same  conditions for the   
dispersion relations as in \ci{LL}.  
Our extension requires the technique  
of holomorphic functions of several complex variables. 
 
In parenthesis we remark that, for the ideal gas,  
 R.Dobrushin and   
Yu.Suhov \ci{DS} first 
realized the importance of a mixing condition  
on the initial measure.  
In \ci{ES} it is replaced by the condition of 
 finite entropy per unit volume thus establishing convergence 
whenever the specific particle number, energy, and 
entropy are finite.  
No such general result seems to be available for the 
harmonic crystal. 
\medskip   
 
We outline our main result and strategy of proof. 
The displacement field $u(x)$ is the deviation of the 
configuration  
of crystal atoms from their equilibrium positions. 
Assuming them to be  
small and expanding the forces to linear order yields 
the discrete  
linear wave equation, 
\be\la{1.1'}      
\ddot u(x,t)  =  -{\sum}_{y\in\Z^d}   
  V(x-y) u(y,t);~~   
\,\,  
u|_{t=0} = u_{0}(x),~~\dot u|_{t=0} = v_{0}(x),
\,\,\,\,x \in\Z^d.     
\ee  
Here   
 $u(x,t)=(u_1(x,t),\dots,u_n(x,t)),   
u_0=(u_{01},\dots,u_{0n})\in\R^n$ and 
correspondingly for $v_0$.   
$ V(x)$ is the interaction (or force) matrix,   
$\Big( V_{kl}(x)\Big),\,\,k,l=1,...,n$.   
The dynamics (\re{1.1'}) is invariant under 
lattice translations.

Let us denote by  
$Y(t)=(Y^0(t),Y^1(t))\equiv   
(u(\cdot,t),\dot u(\cdot,t))$, $Y_0=(Y^0_0,Y^1_0)\equiv   
(u_0(\cdot),v_0(\cdot))$.   
Then (\ref{1.1'}) takes the form of an evolution equation,   
\be\la{CP}   
\dot Y(t)={\cal A}Y(t),\,\,\,t\in\R;\,\,\,\,Y(0)=Y_0.   
\ee   
Formally, this is the  Hamiltonian system since 
\be\la{A}   
{\cal A}Y=  
J 
\left(   
 \begin{array}{cc}   
{\cal V}& 0\\   
0 & 1   
\end{array}\right)Y 
=  
J 
\na H(Y),\,\,\,\,\,\,\,\,\,J= 
\left(   
 \begin{array}{cc}   
0 & 1\\   
- 1 & 0   
\end{array}\right). 
\ee   
Here ${\cal V}$ is a convolution operator with 
the matrix kernel $V$ 
and $H$ is  
the Hamiltonian functional   
 \be\la{ham}  
H(Y)=   
\frac 12\langle v,v\rangle    
+   
\frac 12   
\langle   {\cal V}u, u\rangle, \,\,\,\,\,Y=(u,v),  
\ee  
where the kinetic energy is given by
$\ds\frac 12 \langle v,v\rangle =
 \frac 12 {\sum}_{x\in\Z^d}   
|v(x)|^2$ 
and the potential energy by 
$\ds\frac 12\la{pot} 
\langle{\cal V}  u,u\rangle=\frac 12{\sum}_{x,y\in\Z^d}   
\Big(  V(x-y)u(y), u(x)\Big) 
$,
$\Big(\cdot\, , \cdot \Big)$ being the real scalar product in 
the Euclidean space  
$\R^n$.

We assume that   
the initial datum $Y_0$ is a   
random element of the Hilbert space ${\cal H}_\alpha$   
 of real sequences,   
see Definition \ref{d1.1} below.   
$Y_0$ is distributed according to  the 
 probability measure  $\mu_0$   
of mean zero and satisfying the  
conditions {\bf S1}-{\bf S3} below.   
Given $t\in\R$, denote by $\mu_t$ the probability measure   
for $Y(t)$, the solution to  (\re{CP}) 
with random initial data $Y_0$. 
 We study the   
asymptotics of $\mu_t$ as $t\to\pm\infty$.

The   
correlation  matrices   
of the initial data   
are supposed to be translation-invariant, i.e. for $i,j=0,1$,   
\be\la{1.9'}   
Q^{ij}_0(x,y):= E\Big(Y_0^i(x)\otimes {Y_0^j(y)} \Big)=   
q^{ij}_0(x-y),\,\,\,x,y\in\Z^d,   
\ee   
though our methods require in fact much weaker conditions.   
We also assume that the   
initial mean ``energy'' density  is finite,   
\be\la{med}   
e_0:=E [\vert u_0(x)\vert^2   
 + \vert v_0(x)\vert^2]={\rm tr}\,q_0^{00}(0)+
{\rm tr}\,q_0^{11}(0)   
<\infty,\,\,\,x\in\Z^d.   
\ee   
Finally, it is assumed that the measure $\mu_0$ satisfies a   
mixing   
condition of a Rosenblatt- resp. Ibragimov-Linnik type, 
which means that   
\be\la{mix}   
Y_0(x)\,\,\,\,   {\rm and} \, \, \,\,Y_0(y)   
\,\,\,\,  {\rm are}\,\,\,\, {\rm asymptotically}\,\,\,\, 
{\rm independent}\,\, \,\,   
 {\rm as} \,\, \,\,   
|x-y|\to\infty.   
\ee   
Our main result is the (weak) convergence   
of the measures $\mu_t$ on the Hilbert space 
${\cal H}_\alpha$ with $\al<-d/2$,   
\be\la{1.8i}   
\mu_t \rightharpoondown    
\mu_\infty\,\,\,\mbox{as}\,\,\, t\to \infty.   
\ee
$\mu_{\infty}$
is   
a Gaussian measure on 
${\cal H}_\alpha$.   
A similar convergence result holds for $t\to-\infty$.   
Explicit formulas   
for the correlation functions of  the limit 
measure $\mu_\infty$ 
are given in (\re{Qinfty}) - (\re{QinftyFP}).   
As an application of the results, we show that 
the initial ``white noise''-~correlations
provide the limit measure 
$\mu_\infty$ 
which  coincides with the Gibbs canonical measure 
with the temperature $\sim\! e_0$.  
Respectively, $\mu_\infty$ is close  
to the canonical measure if the initial correlations  
are close to the white noise. 
\medskip

To prove  the convergence (\re{1.8i})
we follow general strategy \ci{BPT,D7, DKKS,DKRS}. 
There are three steps:   
\\   
{\bf I.}   
The family of measures   
 $\mu_t$, $t\geq 0$, is weakly compact in ${\cal H}_\al$
with $\al< -d/2$.\\   
{\bf II.}   
The correlation functions converge to a limit, for $i,j=0,1$   
 \be\la{corf}   
Q_t^{ij}(x,y)= \int Y^i(x)\otimes Y^j(y)\,\mu_t(dY)   
\to Q_\infty^{ij}(x,y)\,\,\,\mbox{as}\,\,\,t\to\infty.   
\ee   
{\bf III.}   
The characteristic functionals converge   
to a Gaussian one, 
\be\la{2.6i}   
 \hat\mu_t(\Psi ):   
 =   \int \exp({i\langle Y,\Psi\rangle})\mu_t(dY)   
\rightarrow \exp\{-\fr{1}{2}{\cal Q}_\infty 
(\Psi ,\Psi)\}\,\,\,\mbox{as}\,\,\,   
t\to\infty.   
 \ee   
Here 
$\Psi=(\Psi^0,\Psi^1)\in{\cal D}=D\oplus D$, 
$D=C_{0}(\Z^d)\otimes \R^n$,
where  $C_{0}(\Z^d)$ denotes the  space of 
the real sequences with finite support,
 $\langle Y,\Psi \rangle  
={\sum}_{i=0,1}{\sum}_{x\in\Z^d}\Big( Y^i(x),
\Psi^i(x)\Big)$ and
${\cal Q}_\infty$ is  
the  
quadratic form 
with the matrix  kernel   
$(Q^{ij}_\infty(x,y))_{i,j=0,1}$, 
\be\la{qpp}   
{\cal Q}_\infty (\Psi, {\Psi})=\sum\limits_{i,j=0,1}~   
\sum\limits_{x,y\in\Z^d}   
\Big(Q_{\infty}^{ij}(x,y),\Psi^i(x)\otimes\Psi^j(y)\Big).   
\ee   
 
Note that (\re{1.1'}) is the translation invariant  
convolution equation
and admits a simple structure in the Fourier space. 
As a consequence, 
Fourier representation plays a central role  
in our proofs of properties {\bf I} and  {\bf II}.  
On the other hand, Fourier transform alone does not 
suffice in proving {\bf III},  
since our main condition (\re{mix}) is stated in the  
coordinate space and its equivalent interpretation in  
Fourier space is obscure.

Property {\bf I} follows by the method \ci{VF}:  
we prove a uniform bound for the covariance of  $\mu_t$  
and refer to the Prokhorov Theorem. 
Property {\bf II} 
is deduced from an analysis of the oscillatory   
 integral representation of the correlation function 
in Fourier space.   
 An important role is attributed to   
 Lemma \re{l4.1}  reflecting   
the properties   
of the  
 Fourier  transformed 
correlation functions  
which is derived from   
the mixing condition. 
To prove {\bf III} we exploit the dispersive properties 
of the dynamics  (\re{1.1'}) in coordinate space.  
The dispersion follows from a stationary phase method 
applied to the oscillatory integral representation of  
the Green's function in Fourier space. 
The dispersion allows us to represent the solution as a sum  
of weakly dependent random variables by the Bernstein-type 
`room-corridor' partition.

Let us explain in more detail the main idea for the proof 
of {\bf III}. 
First let us consider   
the case $n=1$  and the nearest neighbor crystal 
for which  
the   
potential energy has the form 
\be\la{dKG}   
\fr 12
\sum\limits_{x,y\in\Z^d} 
\Big(V(x-y)u(y),u(x)\Big)=   
\fr 12
\sum\limits_{x\in\Z^d}   
(\sum\limits_{i=1}^d|u(x+e_i)-u(x)|^2+   
m^2|u(x)|^2),   
\ee   
where $m\ge 0$ and $e_i=(\de_{i1},\dots \de_{id})$.   
The solution is represented through the Green's function,  
${\cal G} (t,x)$, 
\be\la{solGr}   
Y(x,t)=\sum\limits_{y\in\Z^d}{\cal G} (t,x-y)   
Y_0(y).   
\ee   
The  long-time asymptotics of the  Green's function   
is analyzed by stationary phase method 
based on the dispersion  
relation    
\be\la{omega}   
 \omega(\theta):= \hat V^{1/2}(\theta)=
(2\sum_{j=1}^d(1-\cos\theta_j)+m^2)^{1/2}\,,\,\,\,\,\theta\in \T^d,
\ee 
where $\T^d$ is the real $d$-torus and
$\hat V(\theta)$ stands for 
the Fourier transform of $V(x)$.
The main features of $\om$  for $m> 0$  are   
\be\la{dr}   
i)\,\,\,\om(\theta)\ne 0,\,\,\,\theta\in \T^d\,,\,\,\,\,\,\,\,   
 \mbox{and} \,\,\,\,   
ii) \,\,\mbox{\,\,\,mes}\,{\cal C}=0,   
\ee   
where 
${\cal C}$  is the {\it critical set}   
$     
\{\theta\in \T^d:{\rm det}\,{\rm Hess}\,\om(\theta)=0\}   
$  
and `mes' stands for the Lebesgue measure in $\T^d$.
The Green's function has distinct   
asymptotic behavior in    
 three zones of $(x,t)$-space:  
inside resp. outside the light cone and   
in the `buffer zone',  
 which is a   
small conical neighborhood   
of the   
boundary of the   
light cone.   
The light cone   
is determined by the group velocities   
$\nabla\om(\theta)$ of the phonons,   
and its boundary   
is determined by the group velocities   
$\nabla\om(\theta)$ with ``critical''   
 $\theta\in {\cal C}$,  
since they correspond to the maximal values of  
$|\nabla\om(\theta)|$ with a fixed direction of 
$\nabla\om(\theta)$
(cf. (\re{max})).  
Therefore, the buffer zone   
is determined by the velocities   
$\nabla\om(\theta)$ with the $\theta$ from a small 
neighborhood   
of the critical set   
${\cal C}$.   
The Green's function decays rapidly  outside   
the light cone,   
 as $t^{-d/2}$ inside the light cone   
except for the buffer zone,  
and  
more slowly in the buffer zone, cf. (\re{Gdec}).   
 
Now let us discuss the general case when $n\ge 1$. 
For $n>1$ an additional important feature occurs. 
In this case we have  
$n$ dispersion relations $\om_k(\theta)$, $k=1,...,n$,  
which are the eigenvalues of the  
matrix $\hat V^{1/2}(\theta)$. 
Thus there can be 
``crossing points''   
where two or more dispersion relations $\om_k(\theta)$ 
coincide which implies that they are not differentiable, in general.
In this case the decay   
of the Green's function generally is slower than 
$t^{-d/2}$ everywhere in  
$(x,t)$-space.  
We estimate the decay by the 
stationary phase method, hence we need 
smooth branches of the dispersion relations $\om_k(\theta)$ 
at least locally in $\theta$. We establish the existence of the  
branches outside a set of the Lebesgue measure zero in $\T^d$ 
(see Lemma \re{lc*}). 
For the proof we use 
the advanced variant of  the Weierstrass Preparation Theorem  
from \ci{Ni} and the analytic stratification of analytic sets 
\ci{GR}.

For  $n\ge 1$  
we define the critical set ${\cal C}$ as the subset of $\T^d$ 
which is the union over $k=1,...,n$ of  
all the points $\theta$ either 
 with a nondifferentiable 
$\om_k(\theta)$, or 
with a degenerate Hessian  
of $\om_k(\theta)$, or with $\om_k(\theta)=0$. 
Lemmas \re{lc*}, \re{lc} imply that  
mes~${\cal C}=0$ which plays the central role in all  
proofs in the  paper. 
The critical set is never empty. 
For example, let us 
fix $k=1,...,n$ and
consider the point  
$\theta\in \T^d$ with the maximal group  
velocity $|\nabla\om_k(\theta)|>0$. Then 
${\rm det}\,{\rm Hess}\,\om_k(\theta)=0$ since 
${\rm Hess}\, \om_k(\theta)\, \nabla\om_k(\theta)=0$:
\be\la{max}
\Big({\rm Hess}\, \om_k(\theta)\, \nabla\om_k(\theta)\Big)_i= 
\sum_j\ds\frac{\pa^2\om_k(\theta)}{\pa \theta_i\pa \theta_j}   
\ds\frac{\pa\om_k(\theta)}{\pa \theta_j}=   
\fr 12\ds\frac{\pa}{\pa \theta_i}  
\sum_j  
\Big|\ds\frac{\pa\om_k(\theta)}{\pa \theta_j}\Big|^2=0,
\,\,\,\
i=1,...,d,\,\,
\ee
provided the derivatives exist.
Thus even for 
$d=n=1$   
the {\it uniform in $x\in \R$} decay of the Green's function   
is slower than $t^{-1/2}$   
since $\om''(\theta)$ vanishes in some points.   
To overcome this difficulty,  
in \ci{BPT} it is required   
that $\om'''(\theta)\ne 0$ at points with   
$\om''(\theta)=0$.   
Then the uniform decay of the Green's function   
is $t^{-1/3}$   
which suffices  in the case $d= 1$  
together with an additional assumption   
on the  higher moments of the initial measure. 
In contrast the critical set and   
the slow decay of the Green's function   
do not occur for the   
Klein-Gordon equation analyzed in \cite{D7,DKKS}.

 
For $d,n\ge 1$  Suhov and Shuhov have proved in \ci{SSh}
the convergence of the covariance, (\re{corf}),
for a {\it simple singularity}  of $\om_k(\theta)$ (in Arnold's terminology
\ci{A1}) in the points 
$\theta\in{\cal C}$
with the degenerate Hessian.
However,  
similar detailed analysis of all degenerate points 
for $d,n\ge 1$  
seems to be impossible. We avoid it by  
a novel `cutoff' strategy which allows us to   
cover the general case when 
the Lebesgue measure of the critical set ${\cal C}$ is zero.  
Namely, we choose an $\ve>0$ and 
split the Fourier transform of the solution 
in two components  
$\hat Y(\theta,t)=\hat Y_f(\theta,t)+\hat Y_g(\theta,t)$ where  
$\hat Y_f(\theta,t)=0$ outside the $\ve$-neighborhood of the  
critical set ${\cal C}$ while $\hat Y_g(\theta,t)=0$ inside the  
$\ve/2$-neighborhood of ${\cal C}$. 
First, we use the mixing condition to estimate  
the contribution from the `critical' component $\hat Y_f$:  
we prove that it is small in the mean, i.e. 
its dispersion    
is negligible uniformly in $t\ge 0$, if $\ve>0$ is sufficiently small. 
This  follows from the identity
mes~${\cal C}=0$   
since the Fourier transforms of the initial correlation functions   
are absolutely continuous due to the mixing condition.   
A further step is to develop a Bernstein type argument  
to prove the Gaussian limit for 
the main `noncritical' component 
 $Y_g$. 
We write it in the form (\re{solGr}):   
\be\la{solGrg}   
Y_g(x,t)=\sum\limits_{y\in\Z^d}{\cal G}_g (t,x-y)   
Y_0(y)   
\ee   
where ${\cal G}_g (t,x-y)$ is the `truncated' Green's function 
which is defined 
similarly to $Y_g(x,t)$: its Fourier transform 
 $\hat {\cal G}_g (t,\theta)$ is zero inside  the  
$\ve/2$-neighborhood of ${\cal C}$. Then all the dispersion relations 
$\om_k(\theta)$ are smooth and nondegenerate on the support of  
 $\hat {\cal G}_g (t,\theta)$, hence the truncated Green's function 
has the standard decay 
\be\la{Gdec} 
{\cal G}_g (t,x-y)\le  
\left\{\ba{ll} 
C t^{-d/2},& |y-x|\le c t\\ 
~&\\ 
C_p (|t|+|x-y|+1)^{-p},   & |y-x|\ge c t 
\ea\right. 
\ee 
with some $c>0$ and any $p>0$, 
cf. (\ref{bphi}), (\ref{conp}). Therefore, the representation   
 (\ref{solGrg}) demonstrates that  
for a fixed $x\in\Z^d$, 
the main contribution to $Y_g(x,t)$ comes from the section  
 $B_t(x)=\{y\in\Z^d:\,|y-x|\le c t\}$   
of the light cone at time $t$.   
The ``volume'' of the section   
(i.e. the number of the points $y\in \Z^d\cap B_t(x)$)   
is $|B_t(x)|\sim t^d$.  
Therefore,  
 (\ref{solGrg}) becomes  roughly speaking, 
\be\la{CLTi}   
Y_g(x,t)\sim\fr{\sum\limits_{y\in B_t(x)} Y_0(y)}
{\sqrt{|B_t|}},\,\,\,\,\,t\to\infty.   
\ee   
This implies the Gaussian limit by the  
Ibragimov-Linnik 
Central Limit Theorem   
\ci{IL}, since the random values 
$Y_0(y)$ are weakly dependent because of  the mixing   
condition (\ref{mix}).   
\begin{remarks} i) {\rm Physically, the asymptotics  
(\re{Gdec})  
reflects the isotropic  
phonons propagation of phonons in the noncritical spectrum.  
The isotropy provides a `dynamical mixing' which leads to the  
Gaussian behavior by the statistical mixing condition 
(\re{mix}). 
So the convergence to the statistical equilibrium (\re{1.8i}) 
is  
provided  
by the both kinds of the mixing simultaneously: 
the statistical mixing condition 
(\re{mix}) and the dynamical mixing (\re{Gdec}). 
} 
 
ii) {\rm The degree $-d/2$ in (\re{Gdec}) 
is related to the energy  
conservation  
since the Hamiltonian (\re{ham}) is a quadratic form. 
Roughly speaking,  (\re{Gdec}) means the `energy diffusion',  
and the degree $-d/2$  resembles the 
diffusion kernel. 
} 
\end{remarks} 
 
Finally, let us comment on our conditions concerning 
the interaction matrix  
$V(x)$. 
We assume the conditions   
{\bf E1-E4} below which  
in a similar form appear also in \ci{BPT,LL}.   
{\bf E1} means the 
exponential space-decay of the 
interaction in the crystal.
{\bf E2} resp. {\bf E3}  means that the potential energy is real 
resp. nonnegative.
{\bf E4}  eliminates the constant part of the spectrum  
and ensures   
 that  mes${\cal C}=0$ (cf. (\re{dr})). 
We also introduce a new simple condition {\bf E5}  
for the case $n>1$
which eliminates the {\it discrete} part of the spectrum
for the covariance dynamics.
It can be considerably  
weakened 
to the condition 
{\bf E5'} from Remark \ref{remi-iii} $iii)$.  
For example, the condition {\bf E5'} 
holds for the canonical Gaussian measures  
which are considered in \ci{LL}.
We show that the conditions {\bf E4} and {\bf E5} hold 
for ``almost all'' matrix-functions $V(\cdot)$ with the 
finite range of the interaction.

Furthermore,   
we do not   
require that 
$\om_k(\theta)\ne 0$, $ \theta\in \T^d$:
note that $\om(0)=0$ for  the elastic lattice   
(\ref{omega})  
in the case $m=0$.   
Our results   
hold whenever mes$\{\theta\in \T^d:\om_k(\theta)\!=\!0\}\!=\!0$.  
To cover this case we  impose  the  
new condition {\bf ES} which is roughly speaking  
necessary and sufficient 
for the uniform bounds of the covariance.    
It can be simplified  
to the stronger condition  
\be\la{D'}  
\Vert \hat V^{-1}(\theta)\Vert\in L^1(\T^d)  
\ee  
from \ci{LL}, which holds for the elastic lattice  
(\ref{omega}) if either $n\ge 3$ or $m > 0$.  
The condition (\ref{D'}) is equivalent to {\bf ES} 
for the canonical Gibbs measures considered in \ci{LL}. 
However,   
(\ref{D'}) does not hold  
 in some particular  
interesting cases:  
for instance, for the elastic lattice  
(\ref{omega}) in the case $n=1,2$ and $m=0$,  
as it  is pointed out in \ci{LL}.

The main results of our paper are stated in Section 2:   
Theorem $A$ in Section 2.4, and its application in Section 2.5.4.   
The convergence (\re{corf}) and  
the compactness {\bf I} are established in Section 3,   
  and the convergence (\re{2.6i}) in Sections 4 to 8.   
Section 9  concerns  the ergodicity and the mixing   
properties of the limit measure. 
In Appendix we analyze the crossing points of the dispersion relations.

\setcounter{equation}{0}   
 \section{Main results}   
\subsection{Dynamics}   
  We assume that the initial date $Y_0$   
belongs to the phase space ${\cal H}_\al$,   
 $\al\in\R$,  
defined below.   
 \begin{definition}                 \la{d1.1}   
  $ {\cal H}_\al$   
 is the  Hilbert space   
of pairs $Y=(u(x),v(x))$  of   
 $\R^n$-valued functions  of $x\in\Z^d$ 
 endowed  with  the norm   
 \beqn                              \la{1.5}   
 \Vert Y\Vert^2_{\al}   
 =   
 \sum_{x\in\Z^d}\Big(   
\vert u(x)\vert^2   
 +  \vert v(x)\vert^2\Big)(1+|x|^2)^{\al}   
  <\infty.   
 \eeqn   
  \end{definition}   
We impose the following conditions {\bf E1} - {\bf E5} 
on the matrix   
$ V$.   
\medskip\\   
{\bf E1}   
There exist constants $C,\alpha>0$ such that   
$  
|V_{kl}(z)|\le C e^{-\alpha|z|},\,\,\,\,   
 k,l \in {\rm I}_n:=\{1,...,n\},\,\,\,\,z\in \Z^d.   
$   
\smallskip\\  
Let us denote  by 
$   
\hat V(\theta):=   
\Big(\hat V_{kl}(\theta)   
\Big)_{k,\,l\in {\rm I}_n},   
$   
where   
$   
\hat V_{kl}(\theta)\equiv   
\sum\limits_{z\in\Z^d}V_{kl}(z)e^{iz\theta },$   
$\theta \in \T^d,$   
and  $\T^d$ denotes the $d$-torus   
$\T^d=\R^d/{2\pi \Z^d}$.   
\medskip\\ 
{\bf E2} $ V$ is real and   
symmetric, i.e.   
$V_{lk}(-z)=V_{kl}(z)\in \R$, $k,l \in {\rm I}_n$, 
$z\in \Z^d$.   
\smallskip\\ 
The condition implies that $\hat V(\theta)$ is   
 a real-analytic   
 Hermitian matrix-function in $\theta\in \T^d$.   
\medskip\\   
{\bf E3}   
The matrix   
$\hat V(\theta )$   
is  non-negative definite for each   
 $\theta \in \T^d.$   
\medskip\\   
The condition means that the Eqn  (\re{1.1'}) is 
hyperbolic like  
wave and Klein-Gordon Eqns considered in \ci{DKKS} - \ci{DKS}. 
Let us define the Hermitian  non-negative definite 
 matrix   
\be\la{Omega}   
\Omega(\theta ):=\big(\hat V(\theta )\big)^{1/2}\ge 0   
\ee   
with  the eigenvalues  $\om_k(\theta)\ge 0$,   
$k\in {\rm I}_n$, the dispersion relations.  
For each $\theta\in \T^d$ the Hermitian matrix  
$\Om(\theta)$  has the diagonal form   
 in the  basis  of the orthogonal   
eigenvectors $\{e_k(\theta):k\in  {\rm I}_n \}$:  
\be\la{diag}   
\Om(\theta)=B(\theta)   
\left(\ba{ccc}\om_1(\theta)&\ldots& 0\\   
0&\ddots&0\\   
0&\ldots&\om_n(\theta)   
\ea   
\right)B^*(\theta),   
\ee   
where $B(\theta)$ is a unitary matrix. 
It is well known that the functions  
$\om_k(\theta)$ and  $B(\theta)$ are real-analytic 
outside the set of the `crossing' points $\theta_*$:
$\om_k(\theta_*)=\om_l(\theta_*)$ for some  $l\ne k$.
However, 
generally  the functions  are not smooth 
at the crossing points if $\om_k(\theta)\not\equiv\om_l(\theta)$. 
Therefore, we need the following lemma which we prove in 
Appendix (cf. \ci[Lemma 1.1 ]{Sj}). 
 
\begin{lemma}\la{lc*} 
Let the conditions {\bf E1}, {\bf E2} hold. Then  
there exists  
a closed subset ${\cal C}_*\subset \T^d$ s.t. 
\\ 
i) the Lebesgue measure of ${\cal C}_*$ is zero: 
\be\la{c*} 
{\rm mes}\,{\cal C}_*=0. 
 \ee 
ii) For any point $\Theta\in \T^d\setminus{\cal C}_*$  
there exists a neighborhood 
${\cal O}(\Theta)$ s.t. each dispersion relation 
$\om_k(\theta)$ and the matrix $B(\theta)$  
can be chosen  
as  
the real-analytic functions in 
${\cal O}(\Theta)$. 
\\ 
iii) The eigenvalues $\om_k(\theta)$ have constant multiplicity
in $\T^d\setminus{\cal C}_*$, i.e.
it is possible to enumerate them   
so that we have  for  $\theta\in\T^d\setminus{\cal C}_*$, 
\beqn 
\om_1(\theta)\!\equiv\!\dots\!\equiv\!\om_{r_1}(\theta),\,\,\,   
\om_{r_1+1}(\theta)\!\equiv\!&\dots&\!\equiv\!\om_{r_2}
(\theta),\,\,\,    
\dots \,, \,\,\om_{r_s+1}(\theta)\!\equiv\!\dots\!\equiv\!
\om_{n}(\theta) 
,\la{enum}\\ 
~\nonumber\\ 
\om_{r_\si}(\theta)\!\!\!\!&\!\not\equiv\!&\!\!\!\! 
\om_{r_\nu}(\theta)\,\,\,\,{\rm if}\,\,\,\,\si\ne\nu, 
\,\,\,1\le r_\si,r_\nu\le r_{s+1}:=n. 
\la{enum'} 
\eeqn 
iv) 
The spectral decomposition
holds,
\be\la{spd}
\Om(\theta)=\sum_1^{s+1} \om_{r_\si}
(\theta)\Pi_\si(\theta),
\,\,\,\,\theta\in \T^d\setminus{\cal C}_*,
\ee 
where 
$\Pi_\si(\theta)$  
is the  
orthogonal projection 
in $\R^n$ which is real-analytic function of 
$\theta\in \T^d\setminus{\cal C}_*$.
\end{lemma} 
Below we denote by $\om_k(\theta)$ the local 
real-analytic  
functions from Lemma \re{lc*} $ii)$.
Our next condition is the following: 
\medskip\\ 
{\bf E4}   
$D_k(\theta)\not\equiv 0$,   
$\forall k\in {\rm I}_n $, where   
$D_k(\theta):=   
\det\Big(   
\ds\frac{\pa^2\om_k(\theta)}{\pa \theta_i\pa \theta_j}   
\Big)_{i,j=1}^{d}$, $\theta\in \T^d\setminus{\cal C}_* $.   
\medskip\\  
Let us denote ${\cal C}_0:=\{\theta\in \T^d:\det \hat V(\theta)=0\}$ 
and ${\cal C}_k:=\{\theta\in \T^d\setminus 
{\cal C}_*:\,D_k(\theta)=0\}$, $k\in {\rm I}_n$. 
The following lemma  
is also  
proved in Appendix. 
\begin{lemma}\la{lc} 
Let the conditions {\bf E1} -- {\bf E4} hold. Then  
mes~${\cal C}_k=0$ for $k=0,1,...,n$.   
\end{lemma} 
Our last condition on $V$ is the following:
\medskip\\
{\bf E5}  
For each $k\ne l$  the identity 
$\om_k(\theta)-\om_l(\theta)\!\equiv\!{\rm const}_-$, $\theta\in \T^d$   
does not hold  with const$_-\ne 0$, and
the identity 
$\om_k(\theta)+\om_l(\theta)\!\equiv\!{\rm const}_+$ 
does not hold  with const$_+\ne 0$.   
\medskip
  
This condition holds trivially  in the case $n=1$.
\bigskip  

We show that the conditions {\bf E4} and {\bf E5} 
hold for ``almost all'' functions $V$ 
satisfying the conditions {\bf E1}, {\bf E2}.
More precisely, let us fix an arbitrary $N\ge 1$
and denote by ${\cal R}_N$ the set of
the ``finite range'' interaction
matrices $V$ with $V(x)=0$ for $\max_i |x_i|>N$,
and satisfying the condition {\bf E2}. In Appendix we prove the 
following lemma. 
\begin{lemma}\la{l45}
For any $N\ge 1$
the conditions {\bf E4} and {\bf E5}
hold for  the 
matrix-functions $V$ from an open and dense subset of 
${\cal R}_N$.
\end{lemma}

The following proposition is proved in \ci[p.150]{LL} and \ci[p.128]{BPT}. 
 \begin{pro}    \la{p1.1}   
Let {\bf E1} and {\bf E2} hold,   
and $\al\in\R$.  
Then \\  
i)  
for any  $Y_0 \in {\cal H}_\al$   
 there exists  a unique solution   
$Y(t)\in C(\R, {\cal H}_\al)$   
 to the Cauchy problem (\re{CP}).   
\\   
ii) The operator $U(t):Y_0\mapsto Y(t)$ is continuous in ${\cal H}_\al$.  
\end{pro}   
{\bf Proof}   
Applying the Fourier transform   
to (\ref{CP}),  
we obtain   
\be\la{CPF}   
\dot{\hat Y}(\theta,t)=   
\hat {\cal A}(\theta)\hat Y(\theta,t),\,\,\,t\in\R,   
\,\,\,\,\hat Y(0)=\hat Y_0,  
\ee   
where 
\be\la{hA}   
\hat{\cal A}(\theta)=\left(   
 \begin{array}{cc}   
0 & 1\\   
-\hat V(\theta) & 0   
\end{array}\right),\,\,\,\,\theta\in \T^d.    
\ee   
Note that  $\hat Y(\cdot,t)\in D'(\T^d)$   
 for $t\in\R$.   
On the other hand,  
$\hat V(\theta)$   
is  a smooth function by {\bf E1}.  
Therefore,  
the solution  $\hat Y(\theta,t)$ of (\ref{CPF})   
exists,  is unique and admits the representation   
$\hat Y(\theta,t)=   
\exp\Big(\hat{\cal A}(\theta)t\Big)\hat Y_0(\theta)$. 
It becomes  (\ref{solGr}) in the coordinate space, where the  
  Green's function ${\cal G} (t,z)$ admits the   
Fourier representation   
\be\la{Grcs}  
{\cal G} (t,z):=   
F^{-1}_{\theta\to z}[  
\exp\big(\hat{\cal A}(\theta)t\big)]  
=(2\pi)^{-d}\int\limits_{\T^d}e^{-iz\theta}  
\exp\big(\hat{\cal A}(\theta)t\big)d\theta.  
\ee  
Hence, by the partial integration,   
 ${\cal G} (t,z)\sim|z|^{-p}$   
as $|z|\to\infty$ for any   
$p >0$ and bounded $|t|$  
since $\hat{\cal A}(\theta)$   
is the smooth function of $\theta\in \T^d$.   
Therefore,   
the convolution representation (\ref{solGr})  
implies $Y(t)\in {\cal H}_\al$.   
\hfill$\Box$

\subsection{The convergence to statistical equilibrium}   
   
Let $(\Om,\Si,P)$ be a probability space   
with  expectation $E$   
and  
let  
${\cal B}({\cal H}_\al)$ denote the Borel $\si$-algebra 
in ${\cal H}_\al$.   
We assume that $Y_0=Y_0(\om,\cdot)$ in (\re{CP})   
is a measurable   
random function   
with values in $({\cal H}_\al,\,   
{\cal B}({\cal H}_\al))$.   
In other words, for each $x\in \Z^d$   
the map $\om\mapsto Y_0(\om,x)$   
 is a measurable   
 map   
$\Om\to\R^{2n}$ with respect to the   
(completed)   
$\si$-algebras   
$\Si$ and ${\cal B}(\R^{2n})$.   
Then   
$Y(t)=U(t) Y_0$ is again a measurable  random function   
with values in   
$({\cal H}_\al,{\cal B}({\cal H}_\al))$ owing  to Proposition \re{p1.1}.   
We denote by $\mu_0(dY_0)$ a Borel probability measure   
on ${\cal H}_\al$   
giving   
the distribution of the  $Y_0$.   
Without loss of generality,   
 we assume $(\Om,\Si,P)=   
({\cal H}_\al,{\cal B}({\cal H}_\al),\mu_0)$   
and $Y_0(\om,x)=\om(x)$ for   
$\mu_0(d\om)$-almost all   
$\om\in{\cal H}_\al$ and each $x\in \Z^d$.   
   
\begin{definition}   
$\mu_t$ is a Borel probability measure in   
${\cal H}_\al$   
which gives   
the distribution of $Y(t)$:   
\begin{eqnarray}\la{1.6}   
\mu_t(B) = \mu_0(U(-t)B),\,\,\,\,   
\forall B\in {\cal B}({\cal H}_\al),   
\,\,\,   t\in \R.   
\eeqn   
\end{definition}   
   
Our main goal is to derive   
 the convergence of the measures $\mu_t$ as $t\rightarrow \infty $.   
We establish the weak convergence of  $\mu_t$   
in the Hilbert  spaces ${\cal H}_\al$   
with $\al<-d/2$:   
\be\la{1.8}   
\mu_t\,\buildrel {\hspace{2mm}{\cal H}_\al}\over   
{- \hspace{-2mm} \rightharpoondown }   
\, \mu_\infty   
\quad{\rm as}\quad t\to \infty,   
\ee   
where $\mu_\infty$ is a limit measure on   
the space ${\cal H}_\al$, $\al<-d/2$.   
This means the convergence   
 \be\la{1.8'}   
 \int f(Y)\mu_t(dY)\rightarrow   
 \int f(Y)\mu_\infty(dY),\quad t\to \infty,   
 \ee   
 for any bounded continuous functional $f$   
 on ${\cal H}_\al$.   
   
\begin{definition}   
The correlation functions of the  measure $\mu_t$ are   
 defined by   
\be\la{qd}   
Q_t^{ij}(x,y)= E \Big(Y^i(x,t)\otimes   
 Y^j(y,t)\Big),\,\,\,i,j= 0,1,\,\,\,\mbox{ }\,x,y\in\Z^d,   
\ee   
if the expectations in the RHS are finite.   
Here $Y^i(x,t)$ are the components of the random solution   
$Y(t)=(Y^0(\cdot,t),Y^1(\cdot,t))$.   
\end{definition}   
For a probability  measure $\mu$ on  ${\cal H}_\al$   
we denote by $\hat\mu$   
the characteristic functional (Fourier transform)   
$$   
\hat \mu(\Psi )  =  \int\exp(i\langle Y,\Psi \rangle )\,\mu(dY),\,\,\,   
 \Psi\in  {\cal D}.   
$$   
A measure $\mu$ is called Gaussian (of zero mean) if   
its characteristic functional has the form   
\be\la{G}  
\ds\hat { \mu} (\Psi ) =  \ds \exp\{-\fr12   
 {\cal Q}(\Psi , \Psi )\},\,\,\,\Psi \in {\cal D},   
\ee  
where ${\cal Q}$ is a real nonnegative quadratic form in ${\cal D}$.   
A measure $\mu$ is called   
translation-invariant if   
$   
\mu(T_h B)= \mu(B),\,\,\, B\in{\cal B}({\cal H}_\alpha),   
\,\,\,\, h\in\Z^d,   
$   
where $T_h Y(x)= Y(x-h)$, $x\in\Z^d$.   
   
\subsection{The mixing condition}   
Let $O(r)$ denote the set of all pairs of the subsets   
${\cal A},\>{\cal B}\subset \Z^d$ at distance   
dist$({\cal A},\,{\cal B})\geq r$ and  
$\sigma ({\cal A})$  
be  
the $\sigma $-algebra  in ${\cal H}_\al$ generated by
$Y(x)$   
with $x\in {\cal A}$.   
Define the   
Ibragimov-Linnik mixing coefficient   
of a probability  measure  $\mu_0$ on ${\cal H}_\al$   
by (cf. \ci[Definition 17.2.2]{IL})   
\be\la{ilc}   
\varphi(r):=   
\sup_{({\cal A},{\cal B})\in O(r)} \sup_{   
\ba{c} A\in\si({\cal A}),B\in\si({\cal B})\\ \mu_0(B)>0\ea}   
\fr{| \mu_0(A\cap B) - \mu_0(A)\mu_0(B)|}{ \mu_0(B)}.   
\ee   
\begin{definition}   
 The measure $\mu_0$ satisfies strong, uniform   
Ibragimov-Linnik mixing condition if   
$ 
\varphi(r)\to 0\quad{\rm as}\quad r\to\infty.   
$   
\end{definition}   
Below, we  specify the rate of decay of $\varphi$   
(see condition {\bf S3}).

   
\subsection{Statistical conditions and results}   
We assume that the initial measure $\mu_0$   
satisfies the following conditions {\bf S0--S3}:   
\medskip\\   
{\bf S0}   
$\mu_0$ has zero expectation value,   
$EY_0(x)  \equiv  0$, $x\in\Z^d.$\\   
{\bf S1} $\mu_0$ has translation-invariant correlation matrices,   
 i.e. Eqn   
(\re{1.9'}) holds   
for  $x,y\in\Z^d$.\\   
{\bf S2} $\mu_0$  has a finite mean energy density, i.e.   
Eqn (\re{med}) holds.\\   
{\bf S3}   
 $\mu_0$ satisfies the strong uniform   
Ibragimov-Linnik mixing condition with   
 \be\la{1.12}   
\ov\varphi:=\int\limits_{0}^{+\infty}   
 r^{d-1}\varphi^{1/2}(r)\,dr <\infty.   
 \ee   
We will deduce from {\bf S0} - {\bf S3}  
that $\hat q^{ij}_0\in  C(\T^d)$, $i,j=0,1$
(see Lemma \re{l4.1}). This makes sense of our last condition 
{\bf ES} concerning the initial covariance  
and the matrix $\Om(\theta)$.
We need it only in the case when  
${\cal C}_0\ne\emptyset$ i.e.
det$V(\theta)= 0$ for some points $\theta\in \T^d$:   
\medskip\\  
{\bf ES}
$\Vert \Om^{-i}(\theta)\hat q_0^{ij}(\theta)
\Om^{-j}(\theta)\Vert 
\in L^1(\T^d)$ for $i,j=0,1$.  
\medskip\\ 
This condition follows from {\bf S0} - {\bf S3}
if $i=j=0$ or ${\cal C}_0=\emptyset$.
\medskip\\

Next introduce   
 the correlation matrix of the limit   
measure $\mu_\infty$.   
It 
is translation-invariant (cf. (\re{1.9'})):   
\be\la{Qinfty}   
Q_\infty(x,y)\!=\!\Big(q_\infty^{ij}(x\!-\!y)   
\Big)_{i,j=0,1}.   
\ee   
In the Fourier transform we have locally  
outside the critical set ${\cal C}_*$ 
(see Lemma \re{lc*}) 
\be\la{QinftyF}  
\hat q_\infty^{ij}(\theta)=   
B(\theta)M_{\infty}^{ij}(\theta)B^*(\theta),\,\,\,\,\,\,i,j=0,1, 
\ee  
where   $B(\theta)$ is the smooth 
unitary matrix from Lemma \re{lc*} $ii)$ and 
 $M^{ij}_\infty(\theta)$ is $n\times n$-matrix  with the smooth
entries 
$\Big(M^{ij}_\infty(\theta)\Big)_{kl}=  
\chi_{kl}\Big(   
B^*(\theta) M_0^{ij}(\theta)  
B(\theta)\Big)_{kl}    
$.  
Here we set (see (\re{enum})) 
\be\la{chi}   
\chi_{kl}= 
\left\{ 
\ba{rl} 
1 &{\rm if} \,\,\,\, k,l\in(r_{\si-1}, r_{\si}],\,\,\, 
\si=1,...,s+1, \\ 
 0& {\rm otherwise}  
\ea 
\right. 
\ee 
with $r_0:=0$, $r_{s+1}:=n$,  
and 
\be\la{Pi}   
 M_{0}(\theta):=   
 \frac{1}{2} \left(   
 \ba{cc}   
 \hat q_0^{00}(\theta)+   
\Om^{-1}(\theta)\,\hat q_0^{11}(\theta)\,\Om^{-1}(\theta) &   
\hat q^{01}_0(\theta)-   
\Om^{-1}(\theta)\,\hat q^{10}_0(\theta)\,\Om (\theta) 
\\ ~&\\      
\hat q^{10}_0(\theta)-   
\Om\,\hat q^{01}_0(\theta)\,\Om^{-1}(\theta) &   
 \hat q_0^{11}(\theta)+\Om(\theta)\,\hat  q_0^{00}(\theta)\,
\Om(\theta)   
 \ea     \right).   
\ee   
The local representation 
(\re{QinftyF}) can be expressed globally as 
\be\la{QinftyFP}  
\hat q_\infty^{ij}(\theta)=   
\sum_{\si=1}^{s+1}  
\Pi_\si(\theta)M_0^{ij}(\theta)\Pi_\si(\theta), 
\,\,\,\,\,\,\theta\in \T^d\setminus{\cal C}_*,\,\,\,\,\,\,\,\,\,\,i,j=0,1, 
\ee  
where $\Pi_\si(\theta)$ is the spectral projection 
from (\re{spd}). 
\begin{remark}\la{rem} 
{\rm The condition {\bf ES}  
implies that $( M_0^{ij})_{kl}\in L^1(\T^d)$,  
$k,l\in {\rm I}_n$.  
Therefore,  (\re{QinftyFP}) and  (\re{c*}) 
imply that also $(\hat q^{ij}_\infty)_{kl}\in L^1(\T^d)$,  
$k,l\in {\rm I}_n$. 
}  
\end{remark} 
{\bf Theorem A}   
{\it     Let $d,n\ge 1$,  $\al<-d/2$   
and assume that the conditions {\bf E1--E5}, {\bf S0--S3} hold.
If ${\cal C}_0\ne\emptyset$, then we assume also that 
{\bf ES}  holds.  
 Then \\   
i) the convergence in (\re{1.8}) holds.\\   
ii) The limit measure   
$ \mu_\infty $ is a Gaussian translation-invariant   
measure on ${\cal H}_\al$.\\   
iii) The  characteristic functional of $ \mu_{\infty}$   
is the Gaussian 
\be\la{gau}  
\ds\hat { \mu}_\infty (\Psi ) =   
\exp\{-\fr{1}{2}  {\cal Q}_\infty  (\Psi ,\, \Psi)\},\,\,\,   
\Psi \in {\cal D},   
\ee  
where ${\cal Q}_\infty $ is the quadratic form defined in (\re{qpp}). 
\\   
iv) The measure $\mu_\infty$ is invariant, i.e.   
$[U(t)]^*\mu_\infty=\mu_\infty$, $t\in\R$.

}  
\begin{remarks}\la{remi-iii}   
{\rm  
{\it i)}  
In the case $n=1$, 
the formulas (\re{Pi}), (\re{QinftyFP}) become 
$$   
 \hat q_\infty=   
M_0=   
 \frac{1}{2} \left(   
 \ba{cc}   
 \hat q_0^{00}+   
\om^{-2}\,\hat q_0^{11} &   
\hat q^{01}_0-   
\hat q^{10}_0   \\ ~&\\    
 \hspace{3mm}   
\hat q^{10}_0-   
\hat q^{01}_0 &   
 \hat q_0^{11}+\om^2\,\hat  q_0^{00}   
 \ea     \right).  
$$   
   
{\it ii)}  
 The {\it uniform} Rosenblatt mixing condition   
\ci{Ros} also suffices, together with a higher   
power $>2$  in the bound (\re{med}): there exists $\de >0$ such that   
$$   
E \Big(   
\vert u_0(x)\vert^{2+\de}+\vert v_0(x)\vert^{2+\de}   
\Big)   
<\infty.   
$$   
Then  (\re{1.12}) requires a modification:   
$   
\ds\int_0^{+\infty}  
\ds r^{d-1}\al^{p}(r)dr <\infty$, 
with  
$p=\min(\fr\de{2+\de}, \fr 12)$, 
where $\al(r)$ is the  Rosenblatt   
mixing coefficient  defined   
as in  (\re{ilc}) but without $\mu_0(B)$ in the denominator.   
With these modifications, the statements of Theorem $A$ and their   
proofs remain essentially unchanged.   
  
{\it iii)}  
The arguments with condition {\bf E5}  
in Proposition \ref{p4.1} (see (\ref{4.6'})-(\ref{3.10})  
below)  demonstrate that the condition  
could be considerably weakened.  
Namely, it suffices to assume  
\medskip\\ 
{\bf E5'} If  for some $k\not=l$ we have either
$\om_k(\theta)+\om_l(\theta)\equiv{\rm const}_+\not=0$  
or   
$\om_k(\theta)-\om_l(\theta)\equiv{\rm const}_-\not=0$, 
then  
\be\la{E5'} 
\Big(B^*(\theta)\hat q_0^{ij}(\theta)B(\theta)\Big)_{kl}=0,\,\,\,\,\,  
\theta\in \T^d, \,\, i,j=0,1.  
\ee 
 
}  
\end{remarks}   
 
  
The assertions $i) - iii)$ of Theorem  $A$ follow 
 from Propositions  \re{l2.1} and \re{l2.2}:  
 \begin{pro}\la{l2.1}   
  The family of the measures $\{\mu_t,\,t\in \R\}$   
 is weakly compact in   ${\cal H}_\al$ with any   
 $\al<-d/2$,   
and the bounds hold:   
\be\la{p3.1}   
\sup\limits_{t\ge 0}   
 E \Vert U(t)Y_0\Vert^2_\al <\infty.   
\ee   
 \end{pro}   
 \begin{pro}\la{l2.2}   
  For every $\Psi\in {\cal D}$, the convergence  (\re{2.6i})
 holds.
 \end{pro}   
Proposition \ref{l2.1} ensures the existence of the limit measures 
of the family $\{\mu_t,\,t\in \R\}$, while Proposition 
\ref{l2.2} provides the 
 uniqueness. 
Propositions \ref{l2.1}  and \ref{l2.2} are proved   
in Sections 3 and 4-8, respectively.    
 
Theorem A $iv)$ follows from  
 (\re{1.8})   since the group $U(t)$ is continuous in  
${\cal H}_\al$ by Proposition \re{p1.1} $ii)$. 
  
\subsection{Examples and applications}  
Let us give the examples of the equations  
 (\ref{1.1'})  
and measures $\mu_0$ which  
satisfy all conditions {\bf E1-E5},  
{\bf S0-S3} and {\bf ES}.  
  
\subsubsection{Harmonic crystals}  
All  
 conditions {\bf E1-E5}  
hold for 1D crystal with $n=1$ considered in \ci{BPT}.  
For any $d\ge 1$ and $n=1$  
consider the simple elastic lattice   
corresponding to the quadratic form  
(\ref{dKG}) with $m\ne 0$.  
Then $V(x)=F^{-1}_{\theta\to x}\om^2(\theta)$  
with $\om(\theta)$  defined by (\ref{omega}),  
satisfies {\bf E1-E4}  
with ${\cal C}_*=\emptyset$.  
In these examples the set 
${\cal C}_0$ is  
empty, hence  
the condition {\bf ES} is superfluous. 
Condition {\bf E5} holds trivially since $n=1$.

\subsubsection{Gaussian measures}  
We consider $n=1$ and construct  
Gaussian initial measures $\mu_0$ satisfying 
{\bf S0}--{\bf S3}.  
We will define  $\mu_0$ by  
the correlation functions $q_0^{ij}(x-y)$ which are zero for  
 $i\not= j$, while for $i=0,1$,  
\be\la{S04}  
 \hat q_0^{ii}(\theta) 
:=F_{z\to\theta}q_0^{ii}(z) 
\in L^1(\T^d),\,\,\,\,  
\hat q_0^{ii}(\theta) \ge 0.  
\ee  
Then  by the Minlos theorem \ci{GS} 
there exists a unique Borel 
Gaussian measure  $\mu_0$ on ${\cal H}_\al$, $\al<-d/2$, 
with the correlation functions $q_0^{ij}(x-y)$. 
The measure $\mu_0$  
satisfies {\bf S0-S2}. 
Further, let us provide, in addition to (\re{S04}), that  
\be\la{S5}  
q_0^{ii}(z)=0,\,\,\,|z|\geq r_0.  
\ee  
Then  the mixing condition {\bf S3} follows
with
$\varphi(r)=0$, $r\geq r_0$,  
since  
for  
Gaussian  
random values 
the orthogonality implies the independence.  
For example,   (\re{S04}) and (\re{S5}) hold if we set 
$q_0^{ii}(z)=  
f(z_1)f(z_2)\cdot\dots\cdot f(z_d)$, where  
 $f(z)=  
\nu_0-|z|$ for   
$|z|\le \nu_0$ and $f(z)=0$ for   
$|z|\ge \nu_0$ with  
$\nu_0:=[r_0/\sqrt d]$ (the integer part).  
Then by the direct calculation we obtain  
$  
\hat f(\theta)=\ds\frac{1-\cos{\nu_0\theta}}  
{1-\cos\theta},$  
$\theta\in \T^1,$  
and (\ref{S04}) holds. The measure $\mu_0$  
is nontrivial if  $r_0\ge \sqrt d$: 
otherwise $\nu_0=0$,  
so $q^{ij}_0(z)\equiv 0$, and the measure  $\mu_0(dY_0)$  
is concentrated at the point $Y_0=0$. 
   
\subsubsection{Non-Gaussian measures}  
Let us choose some odd bounded  
nonconstant functions  
 $f^0,\,f^1\in C(\R)$ 
and consider a random function 
 $(Y^0(x),Y^1(x))$  
with  the Gaussian distribution $\mu_0$ from the previous  
example.  
Let us define   
$\mu^*_0$ as the distribution of the random function  
$ (f^0(Y^0(x)),\,  f^1(Y^1(x)))$.  
Then {\bf S0} - {\bf S3} hold for $\mu^*_0$  
with   
corresponding mixing coefficients  
$\varphi^*(r)= 0$ for $r\geq r_0$.  
The measure $\mu^*_0$  
is not Gaussian   if  
 the functions $f^0$, $f^1$ are bounded and nonconstant.  
\subsubsection{From statistical chaos to the Gibbs  measure}  
Let us consider the initial measures 
which satisfy {\bf S0} - {\bf S3}, 
and with the correlation functions 
\be\la{ncc}  
\Big(q_0^{ij}\Big)_{kl}(x-y):= E\Big(Y^i_k(x,0) Y^j_l(y,0)\Big)=
T_i\de_{ij}\de_{kl}\de_{xy}, 
\,\,\,\,i,j=0,1,\,\,\,k,l\in {\rm I}_n, \,\,\,\,x,y\in\Z^d, 
\ee 
where $T_{0,1}\ge 0$. 
These correlations correspond to the 
 ``chaos''
with the zero correlation radius and noncorrelated  
components. 
Such measures exist on ${\cal H}_\al$ with 
$\al<-d/2$ by the Minlos Theorem \ci{GS}: 
for example, the ``white noise'' which is
the  
corresponding Gaussian 
measure.  
Let us consider the crystal satisfying the conditions  
{\bf E1} - {\bf E4} and (\ref{D'}). 
Then also the conditions {\bf E5'}, {\bf ES} hold, 
so  Theorem $A$ is applicable
(see Remark \re{remi-iii} $iii)$)
: it implies the convergence 
(\re{1.8})  
to the Gaussian measure $\mu_\infty$ with the covariance  
(\re{Pi}), (\re{QinftyFP}). 
 
Additionally, let us assume that $T_0=0$ which physically 
means that  
only the initial velocities contribute, and  
initial deviations are adjusted to zero.  
Then the formulas (\re{Pi}), (\re{QinftyFP}) become 
\be\la{gib}   
\hat q_\infty(\theta)= 
 M_{0}(\theta)= 
 \frac{T_1}2 \left(   
 \ba{cc}   
\hat V^{-1}( \theta) &  0  \\  ~&\\     
0 &   
\Big( \de_{kl}\Big)_{k,l\in {\rm I}_n}  
 \ea     \right).   
\ee   
According to (\re{A}), 
this means that the limit measure
$\mu_\infty$ coincides with the  
{\it Gibbs canonical measure} corresponding to the 
temperature $\sim\!T_1$. 
In a more general framework,  the limit measure is close 
to the Gibbs measure if the radius of the initial correlations  
is small in  a suitable scaling limit 
 (cf. \ci[Proposition 4.2]{DKKS}).

\setcounter{equation}{0}   
\section{Convergence of covariance and compactness}   
\subsection{Mixing condition in terms of spectral density}   
The next  
Lemma reflects the mixing property in the 
Fourier transforms   
$\hat q^{ij}_0$   
of initial correlation functions   
$q^{ij}_0$.   
Condition {\bf S2} implies that  $q^{ij}_0(z)$   
is a  bounded  function.   
Therefore, its Fourier transform   
 generally belongs to the Schwartz space   
of tempered distributions.   
\begin{lemma} \la{l4.1}   
Let  the conditions {\bf S0} - {\bf S3} hold.   
Then $\hat q^{ij}_0\in  C(\T^d)$, $i,j=0,1$.   
\end{lemma}   
{\bf Proof} \, It suffices to prove that   
\be\la{4.7} 
q^{ij}_0(z)\in   
l^1(\Z^d).  
\ee 
Conditions {\bf S0} - {\bf S3} imply by 
\ci[Lemma 17.2.3]{IL} (or Lemma \re{il} $i)$ below):   
\be\la{4.9}   
|q^{ij}_0(z)|\le Ce_0\varphi^{1/2}(|z|),~~ z\in\Z^d,   
\ee   
where $e_0$ is defined by (\ref{med}). 
Therefore,  (\ref{1.12}) implies (\re{4.7}): 
$$   
~~~~~~~~~~~~~~~~~~~~~~~~~~~~~~~~~~~~\sum\limits_{z\in\Z^d}   
|q^{ij}_0(z)|\le Ce_0\sum\limits_{z\in\Z^d}   
 \varphi^{1/2}(|z|) <\infty.  ~~~~~~~~~~~~~~~ 
~~~~~~~~~~~~~~~~~~\Box 
$$

\subsection{Oscillatory integral arguments}   
In this section we uniformly estimate and check   
the convergence   
of the correlation matrices of    measures $\mu_t$   
with the help of the Fourier transform.  
The condition 
{\bf S1} and the translation-invariant dynamics  
 (\ref{1.1'}) 
 imply that   
\be\la{4.0}   
Q_t^{ij}(x,y)\equiv   
 \int  Y^i(x)\otimes  Y^j(y)\mu_t(dY)   
=q^{ij}_t(x-y),~~x,y\in\Z^d.   
\ee   
\begin{pro}\la{p4.1}   
i)  
The correlation  
matrices 
$q_t^{ij}(z)$, $i,j=0,1,$ are uniformly bounded   
\be\la{4.40}   
\sup_{t\ge 0}\sup_{z\in\Z^d}   
|q_t^{ij}(z)|<\infty.   
\ee   
ii)  
The correlation  
matrices 
$q_t^{ij}(z)$, $i,j=0,1$,   
converge for each $z\in\Z^d$, and   
\be\la{4.4}   
q_t^{ij}(z)\to q_{\infty}^{ij}(z),~~~~~t\to\infty,   
\ee   
where the functions $q_{\infty}^{ij}(z)$ are   
defined above.   
\end{pro}   
{\bf Proof.}   
For brevity,    
we prove (\ref{4.40}) and (\ref{4.4})   for  $i=j=0$. 
In all other cases the proof of (\ref{4.4}) is similar. 
The solution  to the Cauchy problem (\ref{1.1'})  is   
 $$   
u(x,t)=(2\pi)^{-d}   
\int\limits_{\T^d} e^{-i x\cdot \theta}\Bigl(   
\cos\Om t~ \hat Y_0^0(\theta)+   
\sin \Om t~\Om^{-1} \hat Y_0^1(\theta)   
\Bigr)d\theta,   
$$   
where $\Om\equiv\Omega(\theta)$ is the  
non-negative definite 
Hermitian matrix   
defined by (\ref{Omega}).   
Furthermore, the translation invariance  (\ref{1.9'}) 
implies that 
\be\la{fcor} 
E\Big(\hat Y_0^i(\theta)\otimes  \hat Y_0^j(\theta')\Big)= 
(2\pi)^d\de(\theta+\theta')\hat q_0^{ij}(\theta), 
\,\,\,\,\,i,j=0,1. 
\ee 
Hence,   
\beqn   
q_t^{00}(x\!-\!y)&:=&   
E\Big(u(x,t)\otimes  
 u(y,t)\Big)\nonumber\\  
&\,=&(2\pi)^{-d}\!   
\int\limits_{\T^d}   
e^{-i\theta(x-y)}\Bigl[ \cos \Om t~   
\hat q_0^{00}(\theta)   
\cos \Om t\!+\!\sin \Om t~\Om^{-1}   
\hat q_0^{10}(\theta)\cos \Om t\nonumber\\   
\la{4.6'}   
&&+   
\cos \Om t~   
\hat q_0^{01}(\theta)\Om^{-1}\sin\Om t   
+   
\sin \Om t~\Om^{-1}   
\hat q_0^{11}(\theta)\Om^{-1}\sin\Om t\Bigr]d\theta.   
\eeqn   
Therefore, the bound  
 (\ref{4.40})  with $i=j=0$ follows from Lemma \ref{l4.1}   
or condition {\bf ES} if ${\cal C}_0\ne \emptyset$.  
 
Let us check that the convergence  
 (\ref{4.4}) with $i=j=0$  also follows   
since the oscillatory integrals in (\ref{4.6'})   
 tend to zero. Consider for example   
 the last term in the integrand of (\ref{4.6'}).   
We rewrite  it using  (\ref{diag}), in the form   
\be\la{diaf}   
L_0^{11}(\theta,t):=\sin\Om t~ \Om^{-1}\hat q_0^{11}
(\theta)~\Om^{-1}\sin\Om t  
=B(\theta)\Big(\sin\om_kt~   
A^{11}_{kl}(\theta)\sin\om_l t\Big)_{k,l\in  {\rm I}_n}\,\,
B^*(\theta),  
\ee  
where   
 $A^{11}(\theta):=B^*(\theta)\Om^{-1}\hat q^{11}_0 
(\theta)\Om^{-  
1}B(\theta)$.   
However, at this moment we have to choose certain smooth 
branches 
of the functions $B(\theta)$ and $\om_k(\theta)$ since we 
are going to apply 
the stationary phase arguments which require a smoothness 
in $\theta$. 
To make it correctly, we cut off all singularities. First, 
we define  
the combined {\it critical set} 
\be\la{calC}   
{\cal C}:=\cup_k{\cal C}_k\cup   
{\cal C}_* \cup{\cal C}_0.   
\ee   
Then  Lemmas \re{lc*}, \re{lc} imply the following lemma. 
\begin{lemma}\la{mesF}   
Let conditions {\bf E1} - {\bf E4} hold. Then   
{\rm mes}\,${\cal C}=0$.   
\end{lemma}  
Second, 
fix an $\ve>0$ and choose a finite partition of unity 
\be\la{part}
f(\theta)+g(\theta)=1,\,\,\,\,
g(\theta)=\sum_{m=1}^M g_m(\theta),\,\,\,\,\theta\in \T^d, 
\ee 
where $f,g_m$ are nonnegative functions from 
$C_0^\infty(\T^d)$, the supports of $g_m$ are sufficiently small
and 
\be\la{fge}   
\supp f\subset \{\theta\in \T^d:\,{\rm dist}(\theta,
{\cal C})<\ve\},\,\,\,   
\supp g_m\subset \{\theta\in \T^d:\,{\rm dist}(\theta,
{\cal C})\ge\ve/2\}.   
\ee   
Now   (\ref{diaf}) can be rewritten as  
\beqn \la{fgi}  
&&L_0^{11}(\theta,t)
=f(\theta)L_0^{11}(\theta,t)\nonumber\\ 
&&
\nonumber\\ 
&&
+ 
\frac{1}{2} \sum_{m} g_m(\theta)  
B(\theta) \Big( (\cos(\om_k-\om_l)t -   
\cos(\om_k+\om_l)t) A^{11}_{kl}(\theta)   
\Big)_{k,l\in  {\rm I}_n}\,\,B^*(\theta).   
\eeqn 
By Lemma \re{lc*} and the compactness arguments, 
we can choose the supports of $g_m$ so small that
the eigenvalues $\om_k(\theta)$ 
and the matrix $B(\theta)$  
are  
real-analytic functions inside 
the $\supp g_m$ for every $m$: we do not mark the 
functions by the index 
$m$ to not overburden the notations. 
 
Let us substitute  (\ref{fgi}) into the last term of 
 (\ref{4.6'}) 
and analyze the Fourier 
integrals with  
$f$ and $g_m$ separately. 
The integral with $f$  
converges to zero uniformly in $t\ge 0$, as $\ve\to 0$.  
Indeed, by  Lemma \re{mesF} we have
$$
\Big|\int\limits_{\T^d}   
\!   
e^{-i\theta(x-y)}f(\theta)L_0^{11}(\theta,t)\,d\theta\Big|
\le C
\int\limits_{{\rm dist}(\theta,{\cal C})<\ve}\!\!\!\!\!\!  
\Vert \Om^{-1}(\theta)\hat q_0^{11}(\theta)  
 \Om^{-1}(\theta)\Vert  
\,d\theta\to 0,\,\,\,\,\,\,\ve\to 0
$$
since the 
integrand is
summable by 
Lemma \ref{l4.1} or condition  {\bf ES}  
 if ${\cal C}_0\ne\emptyset$. 
  
Below we will prove the convergence for the  
integrals with $g_m$. 
We will deduce the convergence from the fact that   
the identities
$\om_k(\theta)\pm\om_l(\theta)\equiv$const$_\pm$ 
with the const$_\pm\ne 0$ are impossible
by the condition {\bf E5}. 
Furthermore, 
the oscillatory integrals with 
$\om_k\pm\om_l(\theta)\not\equiv$const vanish as $t\to\infty$.
Hence,
only the integrals with  
$\om_k(\theta)-\om_l(\theta)\equiv 0$  
contribute to the limit 
since  $\om_k(\theta)+\om_l(\theta)\equiv 0$ would imply  
$\om_k(\theta)\equiv\om_l(\theta)\equiv 0$ which 
is impossible by  
 {\bf E4}. 
Similar analysis of the three remaining terms in 
the integrand of 
 (\ref{4.6'}) gives, 
\beqn   
q^{00}_{t}(x-y)=&&\!\!\!\!\!\!\!\!    
(2\pi)^{-d}\int\limits_{\T^d}   
\!   
e^{-i\theta(x-y)}f(\theta)L_0^{11}(\theta,t)\,d\theta
\nonumber\\  
~&&\!\!\!\!\!\!\!\! \nonumber\\ 
&&\!\!\!\!\!\!\!\!+   
(2\pi)^{-d}\sum_{m} \int\limits_{\T^d}   
\!  g_m(\theta)  
e^{-i\theta(x-y)}\Bigl[   
\frac{1}{2} B(\theta)\Big(\chi_{kl}   
 (A_{kl}^{00}(\theta)+A^{11}_{kl}(\theta))   
\Big)_{k,l\in  {\rm I}_n} B^*(\theta)+\dots\Bigr]\,
d\theta\nonumber\\   
~&&\!\!\!\!\!\!\!\!\nonumber\\  
~&&\!\!\!\!\!\!\!\!\nonumber\\  
= 
(2\pi)^{-d}\int\limits_{\T^d}   
\!   
&&\!\!\!\!\!\!\!\!e^{-i\theta(x-y)}f(\theta) 
L_0^{11}(\theta,t)\,
d\theta+ 
(2\pi)^{-d}\int\limits_{\T^d}   
\!  g(\theta)  
e^{-i\theta(x-y)}  
\hat q_\infty^{00}(\theta)\,d\theta  
+\dots \la{3.10} 
\eeqn  
according to the notations (\ref{Qinfty}) - (\ref{Pi}).
Here  
$A^{00}(\theta):=B^*(\theta)\hat q^{00}_0(\theta) B(\theta)$ 
and 
`$\dots$' stands for    
the oscillatory integrals which contain 
 $\cos(\om_k(\theta)\pm\om_l(\theta))t$ 
and $\sin(\om_k(\theta)\pm\om_l(\theta))t$ 
with $\om_k(\theta)\pm\om_l(\theta)\not\equiv$const.

The oscillatory integrals 
converge to zero  
 by  
the Lebesgue-Riemann Theorem  
since all the integrands in `$...$'  are summable and  
$\na(\om_k(\theta)\pm\om_l(\theta))=0$ only on the set 
of the Lebesgue 
measure zero. 
The summability follows from 
Lemma \ref{l4.1} or the condition  {\bf ES}  
 since 
the matrices $B^*(\theta)$ are unitary. 
The zero measure follows similarly to (\re{c*}) 
since $\om_k(\theta)\pm\om_l(\theta)\not\equiv$const.

At last,  
let us prove the convergence (\ref{4.4}) with $i=j=0$. 
From the last line of 
(\ref{3.10}) 
we know that $q^{00}_{t}(x-y)$ is close to the integral with $g$  
if $\ve>0$ is small and $t>0$ is large. 
Therefore, the limit of $q^{00}_{t}(x-y)$ as $t\to\infty$ coincides 
with the limit of the integral as  $\ve\to 0$. 
Finally, this limit coincides with $q^{00}_\infty(x-y)$  since 
$\hat q^{00}_\infty\in L^1(\T^d)$ by Remark \re{rem}. 
\hfill$\Box$   
\subsection{Compactness of measures family}   
{\bf Proof of Proposition \re{l2.1}} The compactness of the 
measures  family  $\{\mu_t,\,t\in \R\}$ will  follow  
from the bounds (\re{p3.1})  
by the Prokhorov Theorem   
 \cite[Lemma II.3.1]{VF}   
using the method  of   
 \ci[Theorem XII.5.2]{VF}  
since the embedding ${\cal H}_\al \subset{\cal H}_\beta$  
is compact if $\al>\beta$.  
 
First, the translation invariance 
 (\ref{4.0}) and Proposition \ref{p4.1} $i)$ imply that 
for $x\in\Z^d$ we have 
 \be   
 e_t:= 
\int [| u_0(x)| ^2 +| v_0(x)|^2]\, \mu_t(dY_0) 
= 
{\rm tr}\,q_t^{00}(0)+{\rm tr}\,q_t^{11}(0)\le \ov e <\infty,   
\,\,\,\, t\ge 0.                \la{3.2}   
 \ee   
 Hence by the definition (\re{1.5}) we get for any $\al<-d/2$:   
$$   
~~~~~~~~~~~~~~~ 
E \Vert U(t)Y_0\Vert^2_\al= e_t   
\sum\limits_{x\in \Z^d}(1+|x|^2)^{\al}=C(\al)e_t\le C(\al)\ov e<\infty,   
\,\,\,\, t\ge 0.~~~~~~~~~~~~~~~~\Box    
$$

\setcounter{equation}{0}   
 \section{   
Duality argument}   
To prove Theorem~A,  
it remains to check Proposition \ref{l2.2}.  
Let us rewrite  (\re{2.6i}) as follows,  
\be\la{*}  
 E\exp \{i\langle Y(t),\Psi\rangle\}  
\to \hat \mu_\infty(\Psi),\,\,\,\,  
t\to\infty.  
\ee    
We will prove it in Sections 5-9.   
 In this section  
 we evaluate $\langle Y(t),\Psi\rangle $    
by using the following duality arguments.   
Remember that $Y_0\in {\cal H}_\al$   
with $\al<-d/2$.   
For  $t\in\R$   
introduce a `formal adjoint' operator   
$U'(t)$   
from  space ${\cal D}$ to ${\cal H}_{-\al}$:   
\be\la{def}   
\langle Y,U'(t)\Psi\rangle =   
\langle U(t)Y,\Psi\rangle ,\,\,\,   
\Psi\in {\cal D},   
\,\,\, Y\in {\cal H}_\al.   
\ee   
Let us denote by $\Phi(\cdot,t)=U'(t)\Psi$.   
Then (\ref{def}) can be rewritten as   
\be\la{defY}   
\langle Y(t),\Psi\rangle =\langle Y_0,\Phi(\cdot,t)\rangle,   
\,\,\,\,t\in\R.   
\ee    
The adjoint group $U'(t)$   
admits  
the following  
convenient description. 
Lemma \re{ldu} below displays that   
the action of group $U'(t)$ coincides with the action   
of  $U(t)$, up to the order of the components.   
\begin{lemma}\la{ldu}   
For $\Psi=(\Psi^0,\Psi^1)\in {\cal D}$ we have   
\be\la{UP}   
\Phi(\cdot,t):=U'(t)\Psi= (\dot\psi(\cdot,t),\psi(\cdot,t)),   
\ee   
where   
 $\psi(x,t)$ is the solution of Eqn (\ref{1.1'})   
 with the initial  
data  
$(u_0,v_0)=(\Psi^1,\Psi^0)$.   
\end{lemma}   
{\bf Proof}   
Differentiating (\ref{def}) in $t$ with   
$Y,\Psi\in {\cal D}$, we obtain  that 
$ 
\langle Y,\dot U'(t)\Psi\rangle =\langle   
\dot U(t)Y,\Psi\rangle.   
$   
The group $U(t)$ has the generator ${\cal A}$ from (\ref{A}).   
The generator of $U'(t)$   is the conjugate operator   
to ${\cal A}$:  
\be\la{A0'}   
{\cal A}'=   
\left( \begin{array}{cc} 0 & -{\cal V} \\   
1 & 0 \end{array}\right).   
\ee   
Hence, the representation (\ref{UP}) holds with   
 $\ddot \psi(x,t)=-\sum\limits_{y\in \Z^d} V(x-y)\psi(y,t)$.   
 \hfill$\Box$\\   
\medskip   
   
The lemma allows us to construct   
the oscillatory integral representation for $\Phi(x,t)$.   
Namely,   
 (\ref{UP}), (\ref{A0'})   imply that   
in the Fourier representation for $\Phi(\cdot,t)=U'(t)\Psi$   
we have   
$$   
\dot{\hat \Phi}(\theta,t)=\hat{\cal A}^* 
(\theta)\hat \Phi(\theta,t),\quad   
\hat \Phi(\theta,t)=\hat{\cal G}^*(t, \theta)   
\hat\Psi(\theta).   
$$   
Here we denote (see (\ref{hA}))   
\be\la{hatA}   
\hat{\cal A}^*(\theta)=   
\left( \begin{array}{cc} 0 & -\hat V(\theta)   \\   
1 & 0 \end{array}\right),   
\quad   
\hat{\cal G}^*(t, \theta)=   
e^{\hat{\cal A}^*(\theta)t}=   
\left(   
 \begin{array}{cc} {\rm cos}~   
\Omega t & -\Omega~{\rm sin}~\Omega t \\   
\Om^{-1}\sin \Omega t   
 & {\rm cos}~\Omega t\end{array} \right)   
\ee   
with $\Om\equiv\Om( \theta)=\Om^*( \theta)$.   
Therefore,   
\be\la{frep'}   
\Phi(x,t)=(2\pi)^{-d}   
\int\limits_{\T^d} e^{-i\theta x}   
\hat{\cal G}^*(t, \theta)\hat\Psi( \theta)~d \theta,   
\,\,\,x\in\Z^d.   
\ee   
Since $f(\theta)+g(\theta)\equiv 1$ by (\re{part}), we can split  
$\Phi$ in two components: 
\beqn\la{frep''}   
\Phi(x,t)&=&(2\pi)^{-d}   
\int\limits_{\T^d} e^{-i\theta x}   
\hat{\cal G}^*(t, \theta)f(\theta)\hat\Psi( \theta)~d \theta+  
(2\pi)^{-d}    
\int\limits_{\T^d} e^{-i\theta x}   
\hat{\cal G}^*(t, \theta)g(\theta)\hat\Psi( \theta)~d \theta 
\nonumber\\ 
~\nonumber\\ 
&=&\Phi_f(x,t)+\Phi_g(x,t),   
\,\,\,x\in\Z^d,   
\eeqn   
where each function $\Phi_f(x,t)$ and $\Phi_g(x,t)$ 
admits the representation of type 
(\ref{UP}). 
By (\ref{fge}), the Fourier spectrum of  $\Phi_f$ 
is concentrated near  
the critical set ${\cal C}$, while the spectrum of $\Phi_g$  
is separated from ${\cal C}$.

 \setcounter{equation}{0}   
 \section{Standard decay in the noncritical spectrum}   
We prove the decay of type (\re{Gdec}) for the 
`noncritical' component  
$\Phi_g$. 
The function  $\Phi_g$ can be expanded  
similarly to (\ref{fgi}), 
 in the form   
\be\la{frepecut}  
\Phi_g(x,t)= 
\sum_{m}  
\sum\limits_{\pm,\,\,k\in  {\rm I}_n }~~   
\int\limits_{\T^d} g_m(\theta)
e^{-i(\theta x\pm\om_k(\theta) t)}   
a^\pm_k(\theta)   
\hat\Psi( \theta)~d \theta.   
\ee   
By Lemma \re{lc*} and the compactness arguments, 
we can choose the eigenvalues $\om_k(\theta)$ 
and the matrices $a^\pm_k(\theta)$  
as  
real-analytic functions inside 
the $\supp g_m$ for every $m$: we do not mark 
the functions by the index 
$m$ to not overburden the notations.

Lemma \re{ldu} means that each component $\Phi_g^i(x,t)$, 
$i=0,1$, 
is a solution to Eqn (\ref{1.1'}).   
To prove (\ref{*}), we analyze the radiative properties   
of $\Phi_g(x,t)$ in all directions.   
For this purpose, we apply the stationary phase method   
to the oscillatory integral (\ref{frepecut})   
along the rays $x=vt$, $t>0$. Then the phase becomes   
$(\theta v\pm\om_k(\theta))t$, and its  
stationary points are the solutions to   
the equations $v=\mp\nabla\om_k(\theta)$.   
We collect all necessary asymptotics in the following lemma 
(cf. (\re{Gdec})). 
\begin{lemma}\la{l5.3}   
For any fixed $\Psi \in {\cal D}$ and   
$g(\theta)\in C_0^\infty (\T^d\setminus {\cal C} )$
the following bounds hold: 
\be\la{bphi}   
\!\!i)~~~~~~~~~~~~~~~~~~~~~~~~~~~~   
\sup\limits_{x\in\Z^d}|\Phi_g(x,t)| \le  C~t^{-d/2}.   
~~~~~~~~~~~~~~~~~~~~~~~~~~~~~~~~~~~~~~~~ 
~~~~~~~   
\ee   
ii) For  any $p>0$ there exist   
$C_p,\ga_g>0$ s.t. 
\be\la{conp}   
~~~~~~~~~~~~~~~~~~~~~~~~~
|\Phi_g(x,t)|\le C_p(|t|+|x|+1)^{-p},\quad\quad   
 |x|\ge \ga_g t.  
~~~~~~~~~~~~~~~~~~~~~   
\ee   
\end{lemma}   
{\bf Proof}  
Consider $\Phi_g(x,t)$  
along each ray   
$x=vt$ with  arbitrary $v\in\R^d$. 
Substituting  
to  
(\ref{frepecut}), we get 
\be\la{freper}   
\Phi_g(vt,t)= 
\sum_{m}  
\sum\limits_{\pm,\,\,k\in  {\rm I}_n }~~   
\int\limits_{\T^d} g_m(\theta)
e^{-i(\theta v\pm\om_k(\theta)) t}   
a^\pm_k(\theta)   
\hat\Psi( \theta)~d \theta.   
\ee   
This is a sum of oscillatory integrals with the phase   
functions $\phi_k^\pm(\theta)=   
\theta v\pm\om_k(\theta)$ 
and the amplitudes $a^\pm_k(\theta) $  
which are real-analytic 
functions of the $\theta$   
inside 
the $\supp g_m$. 
Since $\om_k(\theta)$ is real-analytic, 
 each function $\phi_k^\pm$   
 has no more than a   
 finite number  of  stationary points $\theta\in\supp g_m$,   
solutions to the equation $v=\mp\nabla\om_k(\theta)$.   
The stationary points are non-degenerate for 
$\theta\in\supp g_m$   
by  (\ref{fge}), 
(\ref{calC}) and ${\bf E4}$  
since   
\be\la{Hess}   
{\rm det}\Big(\frac{\pa^2 \phi_k^\pm}
{\pa \theta_i\pa \theta_j}\Big)=   
\pm D_k(\theta)\not= 0,\,\,\,\,\,\theta\in \supp g_m.   
\ee   
At last, $\hat\Psi( \theta)$ is smooth since  $\Psi\in{\cal D}$. 
Therefore, $\Phi_g(vt,t)={\cal O}(t^{-d/2})$   
 according to the standard stationary phase method   
 \ci{F, RS3}.  This implies the bounds (\ref{bphi}) 
in each cone $|x|\le ct$ with any finite $c$. 
 
 Further, denote by   
$\bar v_g:=\max_m\max_{k\in{\rm I}_n} 
\max\limits_{\theta\in \supp g_m}  
|\nabla \om_k(\theta)|.   
$   
Then for $|v|>\bar v_g$ 
the stationary points do not exist on the $\supp g $. 
Hence, the integration by parts as in \ci{RS3} yields 
$\Phi_g(vt,t)={\cal O}(t^{-p})$ for any $p>0$. 
On the other hand, the integration by parts in  
(\ref{frepecut})  
implies similar bound  
$\Phi_g(x,t)={\cal O}\Big(\ds(t/|x|)^l\Big)$ for any $l>0$. 
Therefore, (\ref{conp}) follows with any $\ga_g>\ov v_g$.  
Now the bounds (\ref{bphi}) follow everywhere. 
\hfill$\Box$ 
\bigskip


\setcounter{equation}{0}   
\section{Contribution of  
critical set}   
 
We are going to prove (\ref{*}). Rewrite it using (\ref{defY}): 
\be\la{**}  
 E\exp \{i\langle Y_0,\Phi(\cdot,t)\rangle\}  
-  \hat\mu_\infty(\Psi)\to 0,\,\,\,\,  
t\to\infty.  
\ee    
The splitting (\ref{frep''}) gives 
$\langle Y_0,\Phi(\cdot,t)\rangle =\langle  Y_0,
\Phi_f(\cdot,t)\rangle+   
\langle  Y_0,\Phi_g(\cdot,t)\rangle$. 
Our main argument is that the   
contribution of $\langle  Y_0,\Phi_f(\cdot,t)\rangle$ to   
 (\ref{**}) 
has a small dispersion.   
We will deduce this from Lemmas \re{l4.1}, \re{mesF}.   
At first, let us estimate the difference in (\ref{**}) 
by the triangle inequality:  
\beqn\la{step3}   
&&|E\exp \{i\langle Y_0,\Phi(\cdot,t)\rangle\}  
-\hat\mu_{\infty}(\Psi)| 
\nonumber\\ 
&&~\nonumber\\   
&\le&   
|E\exp\{i \langle U(t)Y_0,\Psi\rangle\}   
- E\exp\{i \langle Y_0,\Phi_g(\cdot,t)\rangle\} |   
+   
|\hat\mu_{\infty}(\Psi_g) -\hat\mu_{\infty}(\Psi) |   
\nonumber\\ 
&&~\nonumber\\    
&& +| E\exp\{i \langle Y_0,\Phi_g(\cdot,t)\rangle\} -   
 \hat\mu_{\infty}(\Psi_g)|   
 = I+II+III,  
\eeqn   
where $\Psi_g:=F^{-1}[g (\theta)\hat\Psi(\theta)]
=\Phi_g(\cdot,0)$. 
Let us consider each of the three terms separately. 
\medskip\\ 
{\bf I.} 
The  first term  $I=I(\ve,t)$    
represents the contribution of the neighborhood of 
the critical set   
$\{\theta\in \T^d:\,{\rm dist}   
(\theta,{\cal C})<\ve\}$   
and   
tends to zero   
 as  $\ve\to 0$ uniformly in $t\ge 0$.   
Namely, by the Cauchy-Schwartz  inequality,  
\beqn\la{I}   
I&=&   
|Ee^{i \langle Y_0,\Phi(\cdot,t)\rangle}   
- Ee^{i \langle Y_0,\Phi_g(\cdot,t)\rangle} | \le   
E|e^{i \langle Y_0,\Phi_f(\cdot,t)\rangle}-1 | \le   
C\Big(E|\langle Y_0,\Phi_f(\cdot,t)\rangle |^2 \Big)^{1/2}.   
\eeqn   
Using the Parseval identity and 
 (\re{frep''}), we get    
\beqn\la{s5.20}   
&&E|\langle Y_0,\Phi_f(\cdot,t)\rangle |^2   
 = (2\pi)^{-2d}E|\langle\hat   
Y_0(\theta),f (\theta)\hat\Phi(\theta,t)\rangle|^2\nonumber\\   
&&~\nonumber\\   
&&= (2\pi)^{-2d} 
\langle 
E\Big(\hat Y_0(\theta)\otimes\ov{\hat Y_0(\theta')}  
\Big), 
\,f (\theta)f (\theta')\,   
\hat {\cal G}^*(t,\theta)\hat\Psi(\theta)\,\otimes   
\ov{\hat{\cal G}^*(t,\theta')}\,   
\ov{\hat\Psi(\theta')}\,\rangle.  
\eeqn 
Now take into account that  
$E\Big(\hat Y_0(\theta)\otimes\overline{\hat Y_0(\theta')}\Big) 
=(2\pi)^d\de(\theta-\theta')\hat q_0(\theta)$ 
similarly to (\re{fcor}). Then   (\re{s5.20}), 
 (\re{hatA}),  (\re{fge}) and the bounds $0\le f (\theta)\le 1$ imply  
$$ 
E|\langle Y_0,\Phi_f(\cdot,t)\rangle |^2 
\le C_1   
\sum\limits_{i,j=0,1}  \,\, 
\int\limits_{{\rm dist}(\theta,{\cal C})<\ve}\!\!\!\!\!\!  
\Vert \Om^{-i}(\theta)\hat q_0^{ij}(\theta)  
 \Om^{-j}(\theta)\Vert  
\,d\theta\to 0,\,\,\,\,  \ve\to 0 
$$  
owing to Lemma \ref{mesF} since the integrand is summable. 
The summability follows from
Lemma \ref{l4.1} 
or 
condition {\bf ES} if ${\cal C}_0\ne\emptyset$.   
\medskip\\ 
{\bf II.}   
The second term $II= II(\ve)$    
tends to zero as $\ve \to 0$.   
Indeed,   
$$  
{\cal Q}_\infty(\Psi_g,\Psi_g)=  
(2\pi)^{-2d}\sum\limits_{i,j=0}^1\,  
\int\limits_{\T^d}  
\Big(\hat q_\infty^{ij}(\theta),  
g(\theta)\hat\Psi^i(\theta) 
\otimes  
g(\theta)\ov {\hat\Psi^j(\theta)}\Big)\,d\theta 
\to 
{\cal Q}_\infty(\Psi,\Psi),  
\,\,\,\,\,\ve\to 0 
$$  
 by  
the Lebesgue Dominated Convergence Theorem  
since $0\le g(\theta)\le 1$ and 
$\hat q_\infty^{ij}\in L^1(\T^d)$  
by Remark \re{rem}. 
Hence for the Gaussian measure  
$\mu_\infty$, we get  
by (\ref{gau})  
$$  
|\hat\mu_\infty(\Psi_g)-  
\hat\mu_\infty(\Psi)|  
=|\exp\{-\fr12{\cal Q}_\infty(\Psi_g,\Psi_g)\}-  
\exp\{-\fr12{\cal Q}_\infty(\Psi,\Psi)\}|\to 0,  
\,\,\,\,\,\ve\to 0.   
$$  
{\bf III.} 
To prove Proposition  \ref{l2.2},   
it remains to check that   
 for any fixed $\ve>0$, we have   
\be\la{II1}   
III(\ve,t)   
=  
| E\exp\{i \langle Y_0,\Phi_g(\cdot,t)\rangle\} -   
 \hat\mu_{\infty}(\Psi_g)|   
\to 0,\,\,\,\,\,\,t\to\infty.   
\ee   
We prove  (\ref{II1}) in Section 8  
using the Bernstein arguments  
of  
the  
next section.

\setcounter{equation}{0}   
\section{Bernstein's `rooms-corridors' partition}   
 
Our proof of (\ref{II1}) is similar to the case   
of the continuous Klein-Gordon equation in $\R^d$   
\ci{DKKS}: 
all the integrals over $\R^d$ become the series over $\Z^d$ etc. 
  Another novelty in the proofs    
is  
the following: 
in the case of the Klein-Gordon equation we have   
$\Phi(x,t)=0$ for $ |x|\ge t+c(\Psi)$, 
while for the discrete crystal we have (\ref{conp}) instead.   

Let us introduce a `room-corridor'  partition of the   
ball $\{x\in\Z^d:~|x|\le \gamma_g t\}$ with $\gamma_g$ from  
(\ref{conp}).   
For $t>0$   
we choose below   
$\De_t,\rho_t\in \N$ (we will specify the  
asymptotical relations between $t$, $\De_t$ and  $\rho_t$).  
Let us set $h_t=\De_t+\rho_t$ and   
\be\la{rom}   
a^j=jh_t,\,\,\,b^j=a^j+\De_t,\,\,\,   
j\in\Z,\,\,\,\,\,\,N_t=[(\gamma_g t)/h_t].   
\ee   
We call the slabs $R_t^j=\{x\in\Z^d: |x|\le N_t h_t,\, a^j\le x_d< b^j\}$   
the `rooms',   
$C_t^j=\{x\in\Z^d: |x|\le  
N_t h_t,\,b^j\le x_d<  a^{j+1}\}$  the `corridors'   
and $L_t=\{x\in\Z^d: |x|> N_t h_t\}$ the 'tails'.   
Here  $x=(x_1,\dots,x_d)$,   
$\De_t$ is the width of a room, and   
$\rho_t$  
is that  
of a corridor. 
Let us denote  by   
 $\chi_t^j$ the indicator of the room $R_t^j$, 
 $\xi_t^j$ that of the corridor $C_t^j$, and  
$\eta_t$ that of the tail $L_t$. 
Then  
\be\la{partB}   
{\sum}_t   
[\chi_t^j(x)+\xi_t^j(x)]+ \eta_t(x)=1,\,\,\,x\in\Z^d,  
\ee   
where the sum ${\sum}_t$ stands for   
$\sum\limits_{j=-N_t}^{N_t-1}$.   
Hence we get the following  Bernstein's type representation:   
\be\la{res}   
\langle Y_0,\Phi_g(\cdot,t)\rangle = {\sum}_t   
[\langle Y_0,\chi_t^j\Phi_g(\cdot,t)\rangle +   
\langle Y_0,\xi_t^j\Phi_g(\cdot,t)\rangle ]+   
\langle Y_0,\eta_t\Phi_g(\cdot,t)\rangle ).   
\ee   
Let us introduce the   
random variables   
 $ r_{t}^j$, $ c_{t}^j$, $l_{t}$ by   
\be\la{100}   
r_{t}^j= \langle Y_0,\chi_t^j\Phi_g(\cdot,t)\rangle,~~   
c_{t}^j= \langle Y_0,\xi_t^j\Phi_g(\cdot,t)\rangle,   
\,\,\,l_{t}= \langle Y_0,\eta_t\Phi_g(\cdot,t)\rangle.   
\ee   
Then  (\ref{res}) becomes 
\be\la{razli}   
\langle Y_0,\Phi_g(\cdot,t)\rangle =   
{\sum}_t   
(r_{t}^j+c_{t}^j)+l_{t}.   
\ee   
\begin{lemma}  \la{l5.1}   
    Let  {\bf S0--S3} hold.   
The following bounds hold for $t>1$:   
\beqn  
E|r^j_{t}|^2&\le&  C(\Psi_g)~\De_t/ t,\,\,\,\forall j,\la{106}\\   
E|c^j_{t}|^2&\le& C(\Psi_g)~\rho_t/ t,\,\,\,\forall j,\la{106''}\\   
E|l_{t}|^2&\le& C_p(\Psi_g)~(1+t)^{-p},\,\,\,\,\forall p>0.\la{106'''}   
\eeqn   
\end{lemma}   
{\bf Proof}   
We discuss  (\ref{106}), and (\ref{106''}), (\ref{106'''})   
can be done in a similar way
(the proof of (\ref{106'''}) additionally uses (\ref{conp})).   
 Express $E|r_t^j|^2$  in the correlation matrices.   
Definition (\ref{100})  implies    
that   
 \be\la{100rq}   
E|r_{t}^j|^2= \langle \chi_t^j(x)\chi_t^j(y)q_0(x-y),   
\Phi_g(x,t)\otimes\Phi_g(y,t)\rangle.   
\ee   
 According to    
(\ref{bphi}), Eqn (\ref{100rq})   
 implies that   
\beqn\la{er}   
E|r_{t}^j|^2&\le&   
Ct^{-d}   
\sum\limits_{x,y}   
\chi_t^j(x)\Vert q_0(x-y)\Vert \nonumber\\  
~ \nonumber\\   
&=&Ct^{-d}\sum\limits_{x}  
\chi_t^j(x) \,\,  
\sum\limits_{z}  
\Vert q_0(z)\Vert\le C \De_t/t,   
\eeqn   
where $\Vert q_0(z)\Vert $ stands for the norm of a matrix   
$\left(q_0^{ij}(z)\right)$.   
Therefore,  (\ref{er})   follows as   
$\Vert q_0(\cdot)\Vert \in l^1(\Z^d)$ by   
(\ref{4.7}).   
\hfill$\Box$

\setcounter{equation}{0}   
\section{Ibragimov-Linnik Central Limit Theorem}  
  
In this section  
 we prove the convergence  (\ref{II1}).  
As was said, we use a version  
of the Central Limit Theorem  
developed by Ibragimov and Linnik \ci{IL}.  
If  ${\cal Q}_{\infty}(\Psi_g,\Psi_g)=0$,  
 (\ref{II1}) is obvious. Indeed,  
$\ds  
|Ee^{i\langle Y_0,\Phi_g(\cdot,t) \rangle}-1|  
\le  
E|\langle Y_0,\Phi_g(\cdot,t) \rangle|\le  
\Big(  
E\langle Y_0,\Phi_g (\cdot,t)\rangle^2  
\Big)^{1/2}  
=  
\Big({\cal Q}_{t}(\Psi_g,\Psi_g)  
\Big)^{1/2}  
$, where   
${\cal Q}_{t}(\Psi_g,\Psi_g)  
\to  
{\cal Q}_{\infty}(\Psi_g,\Psi_g)=0,$ as $t\to\infty.$  
Thus, we may assume that for a given $\Psi\in{\cal D}$,  
\be\la{5.*}  
{\cal Q}_{\infty}(\Psi_g,\Psi_g)\not=0.  
\ee  
Let us choose  $0<\de<1$ and  
\be\la{rN}  
\rho_t\sim t^{1-\delta},  
~~~\De_t\sim\fr t{\log t},~~~~\,\,\,t\to\infty.  
\ee  
\begin{lemma}\la{r}  
The following limit holds true:  
\be\la{7.15'}  
N_t\Bigl(  
\varphi(\rho_t)+\Bigl(  
\frac{\rho_t}{t}\Bigr)^{1/2}\Bigr)  
+  
N_t^2\Bigl(  
\varphi^{1/2}(\rho_t)+\frac{\rho_t}{t}\Bigr)  
\to 0 ,\quad t\to\infty.  
\ee  
\end{lemma}  
{\bf Proof}\,   
The function  
$\varphi(r)$ is non-increasing, hence  by 
(\ref{1.12}),  
\be\la{1111}  
r^{d}\varphi^{1/2}(r)=  
d  
\int\limits_0^r  
s^{d-1}\varphi^{1/2}(r)\,ds\le  
d  
\int\limits_0^r  
s^{d-1}\varphi^{1/2}(s) \,ds\le C\ov\varphi<\infty.  
\ee  
Furthermore, (\ref{rN}) implies  
that   
$h_t=\De_t+\rho_t\sim \ds\frac{t}{\log t}$, $t\to\infty$.  
Therefore, $N_t\sim\ds\frac{t}{h_t}\sim\log t$.  
Then (\ref{7.15'}) follows by (\ref{1111}) and (\ref{rN}).  
\hfill$\Box$\medskip\\  
{\bf Proof of (\ref{II1})}  
By the triangle inequality,  
\beqn  
 |E\exp\{i \langle Y_0,\Phi_g(\cdot,t)\rangle  \}  
-\hat\mu_{\infty}(\Psi_g)|&\le&  
|E\exp\{i \langle Y_0,\Phi_g(\cdot,t)\rangle  \}-  
E\exp\{i{{\sum}}_t r_{t}^j\}|  
\nonumber\\  
&&+|\exp\{-\frac{1}{2}{\sum}_t E|r_{t}^j|^2\} -  
\exp\{-\frac{1}{2} {\cal Q}_{\infty}(\Psi_g, \Psi_g)\}|  
\nonumber\\  
&&+ |E \exp\{i{\sum}_t r_{t}^j\} -  
\exp\{-\frac{1}{2}{\sum}_t E|r_{t}^j|^2\}|\nonumber\\  
&\equiv& I_1+I_2+I_3. \la{4.99}  
\eeqn  
We are going to   show  that all  
the  
summands 
$I_1$, $I_2$, $I_3$  tend to zero  
 as  $t\to\infty$.\\  
{\it Step i)}  
Eqn (\ref{razli}) implies  
\beqn\la{101}  
I_1&=&|E\exp\{i{\sum}_t r^j_{t} \}  
\Big(\exp\{i{\sum}_t c^j_{t}+il_{t}\}-1\Big)|\nonumber\\  
&\le&  
C {\sum}_t E|c^j_t|+E|l_{t}|\le  
C{\sum}_t(E|c^j_t|^2)^{1/2}+(E|l_{t}|^2)^{1/2}.  
\eeqn 
From (\ref{101}), (\ref{106''}), (\ref{106'''}) 
  and (\ref{7.15'}) we obtain that  
\be\la{103}  
I_1\le C_pt^{-p}+ C N_t(\rho_t/t)^{1/2}\to 0,~~t\to \infty.  
\ee  
{\it Step ii)}  
By the triangle inequality,  
\beqn  
I_2&\le& \frac{1}{2}  
|{\sum}_t E|r_t^j|^2-  
 {\cal Q}_{\infty}(\Psi_g, \Psi_g) |  
\le  
 \frac{1}{2}\,  
|{\cal Q}_{t}(\Psi_g, \Psi_g)-{\cal Q}_{\infty}(\Psi_g, \Psi_g)|  
\nonumber\\  
&&+ \frac{1}{2}\, |E\Bigl({\sum}_t r_t^j\Bigr)^2  
-{\sum}_tE|r_t^j|^2| +  
 \frac{1}{2}\, |E\Bigl({\sum}_t r_t^j\Bigr)^2  
-{\cal Q}_{t}(\Psi_g, \Psi_g)|\nonumber\\  
&\equiv& I_{21} +I_{22}+I_{23}\la{104},  
\eeqn  
where ${\cal Q}_{t}$ is  
the  
quadratic form with 
the  matrix kernel $\Big(Q_t^{ij}(x,y)\Big)$.  
(\ref{4.4}) implies that  
 $I_{21}\to 0$.  
As  
for  
$I_{22}$,  we first  
obtain that 
\be\la{i22}  
I_{22}\le \sum_{ \scriptsize\ba{c} j\ne k\\|j|,|k|\le N_t\ea} 
 |Er_t^j r_t^k|.  
\ee  
The next lemma   is the  
corollary of  \ci[Lemma 17.2.3]{IL}. 
\begin {lemma}\la{il}  
 Let ${\cal A},{\cal B}$ be the subsets of $\Z^d$
with the distance {\rm dist}$({\cal A}, {\cal B})\ge r>0$,
and let
$ \xi, \eta$ be random variables  
on the probability space 
$({\cal H}_\al,{\cal B}({\cal H}_\al),\mu_0)$.
Let $ \xi$ be
 measurable with respect to  
the  
$\sigma$-algebra $\sigma({\cal A})$,  
and  $\eta$  with respect to  
the  
$\sigma$-algebra $\sigma({\cal B})$. Then\\  
i)\, 
$\hspace{.5mm}   
|E\xi\eta-E\xi E\eta|\le  
C ab~  
\varphi^{1/2}(r)  
$
if $(E|\xi|^2)^{1/2}\le a$ and $(E|\eta|^2)^{1/2}\le b$.
\\  
ii) 
$  
|E\xi\eta-E\xi E\eta|\le Cab~  
\varphi(r)  \hspace{.5mm}
$  
\,\,\,\,\,\,if $|\xi|\le a$ and $|\eta|\le b$ a.s.

\end{lemma}  
We apply Lemma \re{il} to deduce that  
$I_{22}\to 0$ as $t\to\infty$.  
Note  that  
$r_t^j\!=\!  
\langle Y_0(x),\chi_t^j(x) \Phi_g(\cdot,t)\rangle $  
is measurable  
with respect to the $\sigma$-algebra  $\sigma(R_t^j)$.  
The distance  
between the different rooms $R_t^j$  
is greater or equal to  
$\rho_t$ according to  
 (\ref{rom}).  
Then (\ref{i22}) and (\ref{106}), {\bf S3} imply by  
Lemma \ref{il} $i)$, that  
\be\la{i222}  
I_{22}\le  
C N_t^2\varphi^{1/2}(\rho_t),  
\ee  
which  
tends  
to $0$ as $t\to\infty$ by
 (\ref{7.15'}).  
Finally, it remains to check  
 that $I_{23}\to 0$,  
$t\to\infty$.  
We have   
$$  
{\cal Q}_t(\Psi_g,\Psi_g)  
=E\langle Y_0,\Phi_g(\cdot,t)\rangle^2  
=E\Big({\sum}_t (r_t^j+c_t^j)+l_t\Big)^2  
$$  
according  
to  
(\ref{razli}). 
Therefore, by the  
Cauchy-Schwartz inequality,  
\beqn  
I_{23}&\le&  
 |E\Bigl({\sum}_t r_t^j\Bigr)^2  
- E\Bigl({\sum}_t r_t^j +  
{\sum}_t c_t^j+l_t\Bigr)^2 |\nonumber\\  
& \le&  
C N_t{\sum}_t E |c_t^j|^2  +  
C_1\Bigl(  
E({\sum}_t r_t^j)^2\Bigr)^{1/2}  
\Bigl(  
N_t{\sum}_t E|c_t^j|^2+E |l_t|^2\Bigr)^{1/2}  
+C  E |l_t|^2.\la{107}  
\eeqn  
Then  (\ref{106}), (\ref{i22}) and (\ref{i222})  
imply  
$$  
E({\sum}_t r_t^j)^2\le  
{\sum}_tE|r_t^j|^2 +\sum_{ \scriptsize\ba{c} j\ne k\\|j|,|k|\le N_t\ea} 
 |Er_t^j r_t^k|  
\le  
CN_t\De_t/t+C_1N_t^2\varphi^{1/2}(\rho_t)\le C_2<\infty.  
$$  
Now  
(\ref{106''}),  (\ref{106'''}),  
(\ref{107}) and (\ref{7.15'})  
yield  
\be\la{106'}  
I_{23}\le C_1  N_t^2\rho_t/t+C_2 N_t(\rho_t/t)^{1/2}  
+C_3t^{-p} \to 0,~~t\to \infty.  
\ee  
So,  all  
the  
terms $I_{21}$, $I_{22}$, $I_{23}$ 
in  (\ref{104})  
tend to zero.  
Then  
 (\ref{104}) implies that  
\be\la{108}  
I_2\le  
\frac{1}{2}\,  
|{\sum}_{t}E|r_t^j|^2-  
 {\cal Q}_{\infty}(\Psi_g, \Psi_g)|  
\to 0,~~t\to\infty.  
\ee  
{\it Step iii)}  
It remains to verify that  
\be\la{110}  
I_3=|  
E\exp\{i{\sum}_t r_t^j\}  
-\exp\{-\fr12E\Big({\sum}_t r_t^j\Big)^2\}| \to 0,~~t\to\infty.  
\ee  
Lemma \ref{il}, $ii)$ with $a=b=1$ 
yields:  
\beqn  
&&|E\exp\{i{\sum}_t r_t^j\}-\prod\limits_{-N_t}^{N_t-1}  
E\exp\{i r_t^j\}|  
\nonumber\\  
&\le&  
|E\exp\{ir_t^{-N_t}\}\exp\{i\sum\limits_{-N_t+1}^{N_t-1} r_t^j\}  -  
 E\exp\{ir_t^{-N_t}\}E\exp\{i\sum\limits_{-N_t+1}^{N_t-1} r_t^j\} |  
\nonumber\\  
&&+  
|E\exp\{ir_t^{-N_t}\}E\exp\{i\sum\limits_{-N_t+1}^{N_t-1} r_t^j\}  
-\prod\limits_{-N_t}^{N_t-1}  
E\exp\{i r_t^j\}|  
\nonumber\\  
&\le& C\varphi(\rho_t)+  
|E\exp\{i\sum\limits_{-N_t+1}^{N_t-1} r_t^j\}  
-\prod\limits_{-N_t+1}^{N_t-1}  
E\exp\{i r_t^j\}|.\nonumber  
\eeqn  
Then we  
apply Lemma \ref{il}, $ii)$ recursively 
and get, according to Lemma \ref{r},  
\be\la{7.24'}  
|E\exp\{i{\sum}_{t} r_t^j\}-\prod\limits_{-N_t}^{N_t-1}  
E\exp\{i r_t^j\}|  
\le  
C N_t\varphi(\rho_t)\to 0,\quad t\to\infty.  
\ee  
It remains to check that  
\be\la{110'}  
|\prod\limits_{-N_t}^{N_t-1} E\exp\{ir_t^j\}  
-\exp\{-\fr12{\sum}_{t} E|r_t^j|^2\}| \to 0,~~t\to\infty.  
\ee  
According to the  
standard statement of the  
Lindeberg Central Limit Theorem  
(see, e.g. \ci[Theorem 4.7]{P})  
it suffices to verify the  Lindeberg condition:  
$\forall\de>0$  
$$\frac{1}{\sigma_t}  
{\sum}_t  
 E_{\de\sqrt{\sigma_t}}  
|r_t^j|^2 \to 0,~~t\to\infty.  
$$  
Here  
$  
\sigma_t\equiv {\sum}_t  
E |r^j_t|^2,$  
and $E_a f:= E \Big(X_a f\Big)$,  
where $X_a$ is the indicator of the  
event $|f|>a^2.$  
Note that  
(\ref{108})  
and (\re{5.*}) imply  that  
$  
\sigma_t \to  
{\cal Q}_{\infty}(\Psi_g, \Psi_g)\not= 0,~~t\to\infty.  
$  
Hence it remains to verify that   
$${\sum}_t  
E_{a}  
|r_t^j|^2 \to 0,~~t\to\infty, ~~ \mbox{ for any }\, a>0.  
$$  
This follows   
from the bounds for  
the  
fourth  order moments 
 as in \ci[Section~9]{DKKS}.  
This  completes the proof of  
Proposition \ref{l2.2}.  
\hfill$\Box$  
\setcounter{equation}{0}   
\section{ Ergodicity and mixing  for the limit   
 measures}   
The limit measure $\mu_\infty$ is invariant by Theorem $A$ $iv)$.   
Let $E_\infty$ denote the integral over $\mu_\infty$.   
\begin{theorem}   
Let all assumptions of Theorem A hold for the equation (\re{1.1'}) and the 
initial measure $\mu_0$. Then
$U(t)$ is mixing with respect to   
 the corresponding limit measure $\mu_\infty$, i.e.   
$\forall f,g\in L^2({\cal H}_\al,\mu_\infty)$   
\be\la{3D}   
\lim_{t\to\infty}   
E_\infty f(U(t)Y)g(Y)=   
 E_\infty f(Y)   
E_\infty g(Y).   
\ee   
In particular,   
the group $U(t)$ is ergodic with respect to the measure $\mu_\infty$:   
\be\la{4D}   
\lim_{T\to\infty}   
\frac{1}{T}\int\limits_0^T f(U(t)Y) dt=   
E_\infty f(Y)~~(\mbox{mod }\mu_\infty).   
\ee   
\end{theorem}   
{\bf Proof}\,   
Since $\mu_\infty$ is Gaussian,   
the proof of (\ref{3D})   
reduces to the proof of the following   
convergence:   
$\forall \Psi_1,\Psi_2\in {\cal D}$   
\be\la{5D}   
\lim_{t\to\infty}   
E_\infty \langle U(t)Y,\Psi_1\rangle   
\langle Y,\Psi_2\rangle=0.   
\ee  
 Using the Parseval identity and (\ref{frep''})  
we obtain  similarly to  
(\ref{s5.20}) that 
\beqn\la{6D}   
E_\infty \langle U(t)Y,\Psi_1\rangle   
\langle Y,\Psi_2\rangle &=&  
(2\pi)^{-2d}\int\limits_{\T^d}  
\Big(\hat{\cal G}(t, \theta)\hat q_\infty(\theta ), 
f(\theta)\hat\Psi_1(\theta)\otimes  
\overline{\hat\Psi_2}(\theta)\Big)\,d\theta\nonumber\\ 
~\nonumber\\ 
&+& 
(2\pi)^{-2d}\int\limits_{\T^d}  
\Big(\hat{\cal G}(t, \theta)\hat q_\infty(\theta ), 
g(\theta)\hat\Psi_1(\theta)\otimes  
\overline{\hat\Psi_2}(\theta)\Big)\,d\theta 
\nonumber\\ 
~\nonumber\\ 
&=&I_f(t)+I_g(t). 
\eeqn  
\begin{lemma}\la{lL1}
The uniform bound holds: 
$
\Vert\hat{\cal G}(t, \theta)\hat q_\infty(\theta )\Vert
\le G(\theta)
$, $t\ge 0$, where $G(\theta)\in L^1(\T^d)$.
\end{lemma}
{\bf Proof}
(\ref{hatA}) implies that
\be\la{ap1}
\hat{\cal G}(t, \theta)\hat q_\infty(\theta )=
\left(
\ba{cc}
\cos\Om t&\sin\Om t\\
-\sin\Om t\cdot \Om&\cos\Om t\cdot\Om
\ea\right)
\left(
\ba{cc}
\hat q^{00}_\infty&\hat q^{01}_\infty\\
\Om^{-1}\hat q^{10}_\infty&\Om^{-1}
\hat q^{11}_\infty
\ea\right).
\ee
Therefore,
\be\la{ap0}
\Vert
\hat{\cal G}(t, \theta)\hat q_\infty(\theta )
\Vert\le C\sum\limits_{i,j=0,1}
\Vert\Om^{-i}\hat q^{ij}_\infty(\theta )\Vert.
\ee
It remains to prove that
$\Om^{-i}\hat q^{ij}_\infty(\theta )\in L^1(\T^d)$.
Since
$\hat q_\infty(\theta)\in L^1(\T^d)$ by Remark \re{rem},
it suffices to verify that 
$\Om^{-1}(\theta)\hat q^{1j}_\infty(\theta)\in L^1(\T^d)$,
$j=0,1$.
This also follows from  Remark \re{rem} if ${\cal C}_0=\emptyset$.
Otherwise, we will use the condition {\bf ES}.
Namely,
owing to (\ref{QinftyFP}), we have
\be\la{QinftyFP-1}  
\Om^{-1}(\theta)\hat q_\infty^{ij}(\theta)=   
\sum_{\si=1}^{s+1}  
\Pi_\si(\theta)\Om^{-1}(\theta)M_0^{ij}(\theta)\Pi_\si(\theta) 
\ee  
since $\Om^{-1}(\theta)$ commutes with its spectral 
projection $\Pi_\si(\theta)$.
At last,  (\ref{Pi})
and {\bf ES} imply
$$
~~~~~~~~~~~~~~~~~~
\ba{lclccl}
\Om^{-1}M_0^{10}&=&\ds\frac{1}{2}
\Big(\Om^{-1}\hat q_0^{10}-\hat q_0^{01}\Om^{-1}
\Big)&\in& L^1(\T^d),&\\
~\\
\Om^{-1}M_0^{11}&=&\ds\frac{1}{2}
\Big(\Om^{-1}\hat q_0^{11}+\hat q_0^{00}\Om
\Big)&\in& L^1(\T^d).&
~~~~~~~~~~~~~~~~~~~~~~~~~~~~~
~\Box
\ea
$$
The Lemma \re{lL1} together with (\ref{fge}) and
 Lemma \ref{mesF} imply that 
$\forall \de>0$   
$\exists \ve>0$ such that   
\be\la{7D}   
|I_f(t)|\le \de, 
\,\,\,\,\,\,\,\,t\ge 0.   
\ee   
It remains to study the oscillatory integral $I_g(t)$. 
Rewrite it using (\ref{frepecut}), in the form 
\be\la{sln}  
I_g(t)=\sum_m\sum\limits_{\pm,\,k\in {\rm I}_n}\,\int\limits_{\T^d}   
g_m(\theta)e^{\pm i\om_k(\theta )t}a_k^{\pm}(\theta)   
\Big(\hat q_\infty(\theta ),\hat\Psi_1(\theta)\otimes  
\overline{\hat\Psi_2}(\theta)\Big)\,d\theta.   
\ee 
Here 
all phase functions $\om_k(\theta)$ and  
the amplitudes $a_k^{\pm}(\theta)$ 
are smooth functions in the $\supp g_m$. 
Furthermore, 
 $\na\om_k(\theta)= 0$  
only on the set of the Lebesgue measure zero. 
This follows similarly to (\re{c*}) 
since $\na\om_k(\theta)\not\equiv$const 
by the condition {\bf E4}. 
Hence, 
\be\la{8D}   
I_g(t)\to 0\,\,\,\,\,{\rm as}\,\,\,   
 t\to\infty,   
\ee   
by the Lebesgue-Riemann Theorem since 
 $\hat q_\infty\in L^1(\T^d)$.   
Finally, (\ref{6D})-(\ref{8D})   
imply (\ref{5D}) since $\de>0$ is arbitrary.   
\hfill$\Box$\\   
{\bf Remark}   
Similar result for wave and Klein-Gordon   
 equations   
has been proved in \ci{D7,DK}.   
\setcounter{equation}{0}   
\section{Appendix: Crossing points}   
\subsection{Proof of  Lemmas \re{lc*} and \re{lc}} 
 {\it Step 1}\, 
By the condition {\bf E1}
the matrix $\hat V(\theta )$ is analytic function 
in a connected open (complex) neighborhood ${\cal O}_c(\T^d)$ 
of $\T^d$ in $\T_c^d:=\T^d\oplus i\R^d$. 
Consider the analytic function $d(\theta,\om):=\det (\hat V(\theta )-\om^2)$ 
in ${\cal O}_c(\T^d)\times\C$ and the analytic subset 
defined by 
the equation $d(\theta,\om)=0$ in ${\cal O}_c(\T^d)\times\C$ .  
The subset consists of the points  
$(\theta, \pm \om_k(\theta ))$, $k\in  {\rm I}_n$. 
It is important that 
$d(\theta,\om)\not\equiv 0$ for any fixed $\theta\in 
{\cal O}_c(\T^d)$, 
hence the function $d$ satisfies the {\it Weierstrass condition}  
of  
\ci[Section 2.1.1]{Ni}. Therefore, by 
the Weierstrass Preparation Theorem 
in \ci[Thm 2.1]{Ni},  
there exists a proper analytic  
{\it discriminant 
subset} $\De\subset {\cal O}_c(\T^d)$ s.t.: 
for $\Theta\in  {\cal O}_c(\T^d)\setminus\De$  
there exists a (complex) neighborhood ${\cal O}_c(\Theta)$ 
of $\Theta$ in ${\cal O}_c(\T^d)$
where 
each  
of  
$\om_k(\theta )$ can be chosen  
as  
a holomorphic function. 
More precisely, this is established in 
the proof of \ci[Proposition 2.1]{Ni} which is the main step 
to the proof of the Weierstrass Theorem. 
We set  
${\cal C}_*:=\De\cap \T^d$
and ${\cal O}(\Theta)={\cal O}_c(\Theta)\cap \T^d$ for $\Theta\in \T^d
\setminus {\cal C}_*$. 
Then Lemma \re{lc*} $ii)$ follows for $\om_k(\theta )$.

{\it Step 2}\, 
The identity (\re{c*})  will follow from next general Proposition.
\begin{pro}\la{pms}
Let ${\cal M}$ be a proper analytic subset of 
${\cal O}_c(\T^d)$. Then the Lebesgue measure of 
the intersaction 
$M={\cal M}\cap \T^d$ is zero.
\end{pro}
{\bf Proof}
Let us use the analytic stratification  
of the analytic sets which is constructed 
in  
 \ci[Thm 19 of Chapter II.E and Thm 10 of Chapter III.A]{GR}.  
Namely,  for each $\Theta\in M$ there exists 
a complex neighborhood ${\cal O}_c(\Theta) $ s.t. 
${\cal M}\cap{\cal O}_c(\Theta) =\cup_{0\le\de\le d-1} {\cal M}_\de$, 
where  
each ${\cal M}_\de$ is an analytic submanifold  
of the complex dimension $\de \le d-1$: here we use that   
${\cal M}$ is the proper analytic subset in ${\cal O}_c(\Theta)$. 
Now 
\be\la{repr}
M\cap {\cal O}_c(\Theta)=
\cup_{0\le\de\le d-1} ({\cal M}_\de\cap \T^d).
\ee
\begin{lemma}\la{lMT} 
Let $\Theta\in M$ and
$\de=0,...,d-1$. Then
there exists a (real) neighborhood ${\cal O}(\Theta)$ of $\Theta$
in $\T^d$ s.t. 
the intersection ${\cal M}_\de\cap{\cal O}(\Theta)$
is contained in a  smooth submanifold of  $\T^d$  
of the real dimension $d-1$. 
\end{lemma}  
{\bf Proof} \, We may assume that 
i) ${\cal M}_\de$ is defined by the equations 
$h_j(\theta)=0$, $j=1,...,d-\de$,  
with the holomorphic functions $h_j$
in ${\cal O}_c(\Theta)$,  
and ii) $\na_c\, h_j(\theta)\ne 0$, $\theta\in {\cal O}_c(\Theta)$, 
where $\na_c$ stands for the complex gradient. 
 It is important that $d-\de\ge 1$ so we have  
at least one function $h_1(\theta)$. 
Then  
$h_1(\theta)=f_1(\theta)+ig_1(\theta)$ with the real smooth functions  
$f_1, g_1$, and $f_1(\theta)=g_1(\theta)=0$,  
$\theta\in M_\de\cap{\cal O}_c(\Theta) $. 
However, 
$\na_c\, h_j(\theta)=\na_r \,f_1(\theta)+i\na_r\, g_1(\theta)\ne 0$, 
where $\na_r$ stands for the real gradient.  
Therefore,  either  
$\na_r f_1(\Theta)\ne 0$ or $\na_r g_1(\Theta)\ne 0$. 
\hfill$\Box$ 
\,\bigskip\\
Now Proposition \re{pms} obviously follows.\hfill$\Box$ 

This Proposition implies (\re{c*}) since $\De$ is a proper analytic
subset of 
${\cal O}_c(\T^d)$.
Lemma \re{lc} also follows  from Proposition \re{pms}
since {\bf E4} implies  
that 
$\det \hat V(\theta)\not\equiv 0$ in $\T^d$ and $D_k(\theta)\not\equiv 0$ 
in $\T^d\setminus {\cal C}_*$.

{\it Step 3}\, 
 Lemma \re{lc*} $iii)$ follows 
from the construction in \ci[Section 2.1]{Ni}.
Lemma \re{lc*} $iv)$ follows from (\re{enum'}) since the projection 
$\Pi_\si(\theta)$ can be expressed by the Cauchy  
integral over the contour surrounding the isolated eigenvalue  
$\om_{r_\si}(\theta)$.

{\it Step 4}\, 
It remains to 
 prove Lemma \re{lc*} $ii)$. 
Let ${\cal O}(\Theta)$ denote a small  
real neighborhood of a point $\Theta\in \T^d\setminus {\cal C}_*$
and  $E_\si(\theta)=\Pi_\si(\theta)\R^n$. 
It suffices to construct  
an orthonormal basis  
$\{e_k(\theta): k\in(r_{\si-1}, r_{\si}]\}$ 
in $E_\si(\theta)$  which depends real-analytically on  
$\theta\in{\cal O}(\Theta)$. 
 
Let us  
choose an arbitrary basis  
$\{b_k(\Theta): k\in(r_{\si-1}, r_{\si}]\}$ 
in $E_\si(\Theta)$. Then   
$\Pi_\si(\theta) b_k(\Theta)$ 
 depend real-analytically on  
$\theta\in {\cal O}(\Theta)$, and  
$\{\Pi_\si(\theta) b_k(\Theta): k\in(r_{\si-1}, r_{\si}]\}$ 
is a basis of $E_\si(\theta)$  
for $\theta$ from a reduced neighborhood ${\cal O}'(\Theta)$.  
Finally,  
construct the othonormal basis
$\{e_k(\theta): k\in(r_{\si-1}, r_{\si}]\}$ by the standard Hilbert-Schmidt 
orthogonalization process applied to  
$\{\Pi_\si(\theta) b_k(\Theta): k\in(r_{\si-1}, r_{\si}]\}$ for each 
$\theta\in{\cal O}'(\Theta)$. 
\hfill$\Box$ 
   
 \begin{remark}\la{rcr} 
{\rm Lemma \re{lc*} $iii)$ also follows from  
\ci[Section 2.1]{Ni} since 
the enumeration (\re{enum}), (\re{enum'}) 
corresponds to the factorization of type 
\ci[(2.5)]{Ni} for the 
function $d(\theta,\om)$,  
into the product of the irreducible factors, 
with the multiplicities $r_\si-r_{\si-1}$, which is constructed in 
\ci[Thm 2.1]{Ni}.} 
\end{remark} 

\subsection{Proof of Lemma \re{l45}}
 {\it Step 1}\, 
Let us fix
arbitrary $k,l\in  {\rm I}_n$
and consider 
$\om_k( \theta)$
as the functions of $V\in{\cal R}_N$
and of $\theta\in\T^d$.
It suffices to prove that 
$D_k( \theta)$ and 
$\na(\om_k( \theta)\pm\om_l( \theta))$ are analytic and 
are 
not zero
in an open dense subset in ${\cal R}_N\times\T^d$.

Let us consider $V_{k'l'}(x)$, $k',l'\in  {\rm I}_n$, $|x_i|\le N$,
as the coordinates of the matrix-function $V$
in the region ${\cal R}_N$. Condition {\bf E2} allows us to 
consider $V_{k'l'}(x)$
as independent real variables for any $k',l'\in  {\rm I}_n$ and 
the points $x$ with either
 $x_1> 0$, or $x_1=0$ and $x_2 > 0$, or $x_1=x_2=0$ and
$x_3 > 0$, etc. Let us identify ${\cal R}_N$ with 
corresponding  range $\R^M$ of the independent real variables $V_{k'l'}(x)$.

 {\it Step 2}\,
Consider $\om_k(\theta)$ as the 
functions of $\{V_{k'l'}(x)\}$ and $\theta$
 in $\C^M\times\T_c^d$.
As above, each $\om_k(\theta)$ can be chosen as  
a holomorphic function outside a proper analytic discriminant subset
$\De\subset\C^M\times\T_c^d $. Lemma \re{lMT}
implies that the region 
$O:=(\R^M\times\T^d)\setminus\De $ 
is an open dense subset in
$\R^M\times\T^d$. 
Therefore, it suffices to prove that the functions
$D_k$ and $\na(\om_k\pm\om_l)$ are not identically zero in each 
connected open component of $O$.
However,  the region of analyticity  
${\cal O}:=(\C^M\times\T_c^d)\setminus \De$ is connected.
Hence, 
it remains to construct
a point of
$\C^M\times\T_c^d$ s.t.
the functions $D_k$ and $\na(\om_k\pm\om_l)$
are holomorphic and non identically zero
in a neighborhood of the point.
It is easy to construct  
such point for any $n\ge 1$:
we can choose an arbitrary $\theta\in\T^d$ and 
the nearest neighbor crystal (\ref{dKG}) repeated
$n$ times with distinct masses $m_k$, $k\in {\rm I}_n$.
\hfill$\Box$

   
   \end{document}